\documentclass[12pt]{article}
\usepackage{fancyhdr}
\usepackage{amsmath, amsthm, amsfonts, amssymb}
\usepackage{mathrsfs} 
\usepackage{dsfont}
\usepackage[T1]{fontenc}
\usepackage{mathpazo}
\usepackage{setspace}
\usepackage{epsfig}
\usepackage{latexsym}
\usepackage{color}
\usepackage{graphicx}
\usepackage{nicefrac}
\usepackage[latin1]{inputenc}
\usepackage{slashed}
\usepackage{multirow}
\usepackage{comment}
\usepackage{soul}
\usepackage{cite}
\usepackage{dsfont}
\usepackage[all,cmtip]{xy}
\xyoption{arc}
\usepackage{graphics}
\usepackage[colorlinks=true, pdfstartview=FitV, linkcolor=black, citecolor=blue, urlcolor=blue]{hyperref}
\usepackage{esint}

\usepackage{ifpdf}
\ifpdf
\usepackage{epstopdf,hyperref}
\else
\usepackage[hypertex]{hyperref}
\fi

\usepackage[toc]{appendix}

\def\hybrid{\topmargin -20pt    \oddsidemargin 0pt
        \headheight 0pt \headsep 0pt
        \textwidth 6.25in       
        \textheight 9.25in       
        \marginparwidth .875in
        \parskip 5pt plus 1pt   \jot = 1.5ex}

\hybrid

\def\baselinestretch{1.2}

\catcode`\@=11

\def\marginnote#1{}
\newcount\hour
\newcount\minute
\newtoks\amorpm
\hour=\time\divide\hour by60
\minute=\time{\multiply\hour by60 \global\advance\minute by-\hour}
\edef\standardtime{{\ifnum\hour<12 \global\amorpm={am}%
        \else\global\amorpm={pm}\advance\hour by-12 \fi
        \ifnum\hour=0 \hour=12 \fi
        \number\hour:\ifnum\minute<10 0\fi\number\minute\the\amorpm}}
\edef\militarytime{\number\hour:\ifnum\minute<10 0\fi\number\minute}

\def\draftlabel#1{{\@bsphack\if@filesw {\let\thepage\relax
   \xdef\@gtempa{\write\@auxout{\string
      \newlabel{#1}{{\@currentlabel}{\thepage}}}}}\@gtempa
   \if@nobreak \ifvmode\nobreak\fi\fi\fi\@esphack}
        \gdef\@eqnlabel{#1}}
\def\@eqnlabel{}
\def\@vacuum{}
\def\draftmarginnote#1{\marginpar{\raggedright\scriptsize\tt#1}}

\def\draft{\oddsidemargin -.5truein
        \def\@oddfoot{\sl preliminary draft \hfil
        \rm\thepage\hfil\sl\today\quad\militarytime}
        \let\@evenfoot\@oddfoot \overfullrule 3pt
        \let\label=\draftlabel
        \let\marginnote=\draftmarginnote
   \def\@eqnnum{(\theequation)\rlap{\kern\marginparsep\tt\@eqnlabel}%
\global\let\@eqnlabel\@vacuum}  }

\def\preprint{\twocolumn\sloppy\flushbottom\parindent 2em
        \leftmargini 2em\leftmarginv .5em\leftmarginvi .5em
        \oddsidemargin -.5in    \evensidemargin -.5in
        \columnsep .4in \footheight 0pt
        \textwidth 10.in        \topmargin  -.4in
        \headheight 12pt \topskip .4in
        \textheight 6.9in \footskip 0pt
        \def\@oddhead{\thepage\hfil\addtocounter{page}{1}\thepage}
        \let\@evenhead\@oddhead \def\@oddfoot{} \def\@evenfoot{} }

\def\numberbysection{\@addtoreset{equation}{section}
        \def\theequation{\thesection.\arabic{equation}}}

\def\underline#1{\relax\ifmmode\@@underline#1\else
        $\@@underline{\hbox{#1}}$\relax\fi}

\def\titlepage{\@restonecolfalse\if@twocolumn\@restonecoltrue\onecolumn
     \else \newpage \fi \thispagestyle{empty}\c@page\z@
        \def\thefootnote{\fnsymbol{footnote}} }

\def\endtitlepage{\if@restonecol\twocolumn \else \newpage \fi
        \def\thefootnote{\arabic{footnote}}
        \setcounter{footnote}{0}}  

\catcode`@=12
\relax

\def\figcap{\section*{Figure Captions\markboth
        {FIGURECAPTIONS}{FIGURECAPTIONS}}\list
        {Figure \arabic{enumi}:\hfill}{\settowidth\labelwidth{Figure
999:}
        \leftmargin\labelwidth
        \advance\leftmargin\labelsep\usecounter{enumi}}}
 \relax
\def\tablecap{\section*{Table Captions\markboth
        {TABLECAPTIONS}{TABLECAPTIONS}}\list
        {Table \arabic{enumi}:\hfill}{\settowidth\labelwidth{Table
999:}
        \leftmargin\labelwidth
        \advance\leftmargin\labelsep\usecounter{enumi}}}
 \relax
\def\reflist{\section*{References\markboth
        {REFLIST}{REFLIST}}\list
        {[\arabic{enumi}]\hfill}{\settowidth\labelwidth{[999]}
        \leftmargin\labelwidth
        \advance\leftmargin\labelsep\usecounter{enumi}}}
 \relax
\makeatletter
\newcounter{pubctr}
\def\publist{\@ifnextchar[{\@publist}{\@@publist}}
\def\@publist[#1]{\list
        {[\arabic{pubctr}]\hfill}{\settowidth\labelwidth{[999]}
        \leftmargin\labelwidth
        \advance\leftmargin\labelsep
        \@nmbrlisttrue\def\@listctr{pubctr}
        \setcounter{pubctr}{#1}\addtocounter{pubctr}{-1}}}
\def\@@publist{\list
        {[\arabic{pubctr}]\hfill}{\settowidth\labelwidth{[999]}
        \leftmargin\labelwidth
        \advance\leftmargin\labelsep
        \@nmbrlisttrue\def\@listctr{pubctr}}}
 \relax
\makeatother
\newskip\humongous \humongous=0pt plus 1000pt minus 1000pt

\newif\ifdtup

\relax

\def\ba{\begin{eqnarray}}
\def\ea{\end{eqnarray}}

\def\a{\alpha}

\def\b{\beta}

\def\g{\gamma}

\def\d{\delta}
\def\D{\Delta}

\def\th{\theta}

\def\m{\mu}
\def\n{\nu}

\def\l{\lambda}

\def\s{\sigma}

\def\Hc{{\cal H}}

\def\({\left(} \def\){\right)}

\def\no{\noindent}

\def\qq{\qquad}

\def\IR{\relax{\rm I\kern-.18em R}}

\def \ov {\over}

\def\IR{\relax{\rm I\kern-.18em R}}
\def\IL{\relax{\rm I\kern-.18em L}}

\def\inv{^{\raise.15ex\hbox{${\scriptscriptstyle -}$}\kern-.05em 1}}

\def\dd{\mathrm{d}}
\def\ii{\mathrm{i}}
\DeclareMathOperator{\Img} {\mathrm{Im}}

\begin{document}

\renewcommand{\theequation}{\thesection.\arabic{equation}}
\csname @addtoreset\endcsname{equation}{section}

\newcommand{\mt}[1]       {\textrm{\tiny #1}}
\newcommand\be            {\begin{equation}}
\newcommand\bea           {\begin{equation}\begin{array}l\displaystyle}
\newcommand\bearll        {\begin{array}{ll}\displaystyle}
\newcommand\ee            {\end{equation}}
\newcommand\eear          {\end{array}}
\newcommand\scr           {\scriptstyle}

\newcommand{\beq}{\begin{equation}}
\newcommand{\eeq}[1]{\label{#1}\end{equation}}
\newcommand{\ber}{\begin{equation}}
\newcommand{\eer}[1]{\label{#1}\end{equation}}
\newcommand{\eqn}[1]{(\ref{#1})}
\begin{titlepage}
\begin{center}


${}$
\vskip .2 in

{\large\bf Non-Abelian T-duality and Modular Invariance}\\

\vskip 3em

{\bf Benjo Fraser},$^1$ {\bf Dimitrios Manolopoulos$^{2,3}$}\ and\ {\bf Konstantinos Sfetsos$^{2}$}
\vskip 0.1in

{\em
${}^1$CUniverse Group, \\
Department of Physics,\\
Faculty of Science, Chulalongkorn University,\\
Bangkok 10330, Thailand\\
}

\vskip 0.1in
 {\em
${}^2$Department of Nuclear and Particle Physics,\\
Faculty of Physics, National and Kapodistrian University of Athens,\\
Athens 15784, Greece\\
}

\vskip 0.1in
{\em
${}^3$Computation-based Science and Technology Research Center,\\
Cyprus Institute, 20 Kavafi Str., Nicosia 2121, Cyprus\\
}

\vskip 0.1in

{\footnotesize \texttt {benjojazz@gmail.com, d.manolopoulos@icloud.com, ksfetsos@phys.uoa.gr}}

\vskip 1em
\vskip 3em

\end{center}

\centerline{\bf Abstract}

\no
Two-dimensional $\s$-models corresponding to coset CFTs of the type $ (\hat{\mathfrak{g}}_k\oplus \hat{\mathfrak{h}}_\ell )/  \hat{\mathfrak{h}}_{k+\ell}$ admit a zoom-in limit involving sending one of the levels, say $\ell$, to infinity. The result is the non-Abelian T-dual of the WZW model for the algebra $\hat{\mathfrak{g}}_k$ with respect to the vector action of the subalgebra $\mathfrak{h}$ of $ \mathfrak{g}$. We examine modular invariant partition functions in this context.
Focusing on the case with $\mathfrak{g}=\mathfrak{h}=\mathfrak{su}(2)$ we apply the above limit to the branching functions and modular invariant partition function of the coset CFT, which as a whole is a delicate procedure.
Our main concrete result is that such a limit is well defined and the resulting partition
function is modular invariant.

\vskip .4in
\noindent
\end{titlepage}
\vfill
\eject

\newpage

\tableofcontents

\noindent

\def\baselinestretch{1.2}
\baselineskip 20 pt
\noindent


\setcounter{equation}{0}

\section{Introduction}

In this paper we consider quantum aspects of non-Abelian T-duality in Wess-Zumino-Witten (WZW) \cite{Witten:1983ar,Gepner:1986wi} models. These are two-dimensional sigma models with a group manifold as target space. Conformal invariance is exact in the curvature expansion, so they are useful examples for study of string theory compactifications.

Duality in these theories received attention for a number of years \cite{Rocek:1991ps,Giveon:1993ai,Sfetsos:1994vz,Alvarez:1994np,Gaberdiel:1995mx}. Some of the conclusions drawn at that time were the following. In the Abelian case it is possible to specify periodicities for the dual coordinates such that the duality holds at the quantum level - in particular the spectra of the two theories
 match. However, in the non-Abelian case this does not seem possible: some coordinates in the dual geometry must be left non-compact, and so non-Abelian `duality' is actually a mapping between different CFTs. There was a generalization of these ideas to Poisson-Lie Duality \cite{Klimcik:1995ux,Alekseev:1995ym,Klimcik:1996nq,Klimcik:1996hp}.
 While progress was made, the quantum picture of the Non Abelian T-duality was never completely elucidated.

Here we reconsider the case of the $\hat{\mathfrak{su}}(2)_k$ WZW model on the torus, where a vital role in the consistency of the theory is played by modular invariance. We explicitly solve for the spectrum of the dual quantum theory and compute the partition function. We also consider modular transformations and fusion rules.

The tool we use to do this is the coset construction \cite{Goddard:1984vk}\cite[Chap. 18]{DiFrancesco:1997nk}.
As shown in \cite{Sfetsos:1994vz}, the classical action for the non-Abelian T-dual of the WZW model for $SU(2)$ with respect to the vector $SU(2)$ isometry can be obtained as the limit of the coset model
\begin{align}\label{eq:coset su2 diagonal}
\frac{\hat{\mathfrak{su}}(2)_k\oplus \hat{\mathfrak{su}}(2)_{\ell}}{\hat{\mathfrak{su}}(2)_{k+\ell}},
\end{align}
in which $\ell\rightarrow\infty$, and we must also zoom in close to the identity in the corresponding group geometry, parametrizing the second group element as $g_2=\mathds{1}+\mathrm{i}\, v/\ell$. This limit allows for very large quantum numbers for the algebra $\hat{\mathfrak{su}}(2)_{\ell}$.\footnote{More generally, in the limit $\ell\to \infty$, the $\s$-model corresponding to the coset CFT
$
(\hat{\mathfrak{g}}_k\oplus \hat{\mathfrak{h}}_\ell)/\hat{\mathfrak{h}}_{k+\ell},
$
is the non-Abelian T-dual of the WZW model for $G$ with respect to the vector action
of the subgroup $H$ \cite{Sfetsos:1994vz}.}
The group theoretic coset specifies only the chiral algebra of the theory, and it was our motivation in this paper to understand precisely which sectors of left and right moving quantum numbers can combine to give a modular invariant partition function as $\ell\rightarrow\infty$. Our answer will be that there is a consistent subset of primary states, closed under modular transformations, which implement the limit. The defining feature of these states is that they have finite non-zero energy.

Our main result is that the dual partition function is equal to that for a product theory of parafermions and an uncompactified boson, $(\hat{\mathfrak{su}}(2)_k/\hat{\mathfrak{u}}(1)_k)\times \mathbb{R}$. Furthermore we find that by taking two inequivalent limits in target space, we can associate this partition function with two \textit{different} geometries. This is suggestive of a duality between these two backgrounds, which is not the original non-Abelian T-duality, but something else, a kind of large-level equivalence. This idea has been put forward  in a previous work \cite{Polychronakos:2010fg, Polychronakos:2010hd} where the spectrum of scalar Laplacian operator was considered, but here it is made explicit and verified by considering the whole tower of string states. As far as we know, both this conjectured equivalence and the result for the CFT partition function we find here are new.

It has long been known \cite{Rocek:1991ps,Giveon:1993ai} that both Abelian and non-Abelian T-duality \cite{Alvarez:1994dn} can be thought of on the string worldsheet as a continuous orbifold, i.e. an orbifold by a continuous group. While we have not checked explicitly, the form of the limiting partition function we find is highly suggestive of the same interpretation. Indeed,
in previous works \cite{Gaberdiel:2011aa,Restuccia:2013tba,Gaberdiel:2014vca,Fredenhagen:2014kia} similar limits were taken and there were interpreted as continuous orbifolds. The most well-known example is the so-called Runkel-Watts theory \cite{Runkel:2002yb,Roggenkamp:2003qp}, corresponding to the $c\rightarrow 1$ limit of the Virasoro minimal models. This is a special case of our results for $k=1$.

The paper is organized as follows: As a warm-up in section \ref{sec:wzw} we consider the $\hat{\mathfrak{su}}(2)_k$ WZW model, which is the model before dualization, and its large-$k$ limit. While this is not the limit we are directly interested in, it will serve to illustrate some relevant points in a simpler context. In section \ref{sec:Coset th} we consider the large-level limit of the coset theory, and discuss a consistent set of states to preserve modular invariance. We also give a spacetime interpretation of our results. In section \ref{sec:abelian}, for the sake of comparison, we treat the well studied case of Abelian duality using our methods. Finally section \ref{sec:discussion} ends with a discussion of our results.

\section{Large-level limit of the \texorpdfstring{$\hat{\mathfrak{su}}(2)_k$}{su(2)} WZW model}
\label{sec:wzw}
Our approach to modular invariants for backgrounds related to non-Abelian T-duality
was first established in \cite{Sfetsos:1994vz} by taking a limit
in gauged WZW models (corresponding to coset CFTs), involving as a basic ingredient large level.
We would like to see how this limit manifests itself at the level of the partition
function of the theory. As we will see this is technically quite involved.
Therefore, in order to introduce our approach and method, we start with
a different (albeit related and much simpler) limit in WZW models.

\subsection{The background}
We focus our discussion on the $\hat{\mathfrak{su}}(2)_k$ WZW model.
In a standard parametrization of the group element
\be
g = \mathrm{e}^{\frac{\mathrm{i}}{2}(\phi_1-\phi_2)\sigma_3}
\mathrm{e}^{\mathrm{i}\omega\sigma_2}\mathrm{e}^{\frac{\mathrm{i}}{2}(\phi_1+\phi_2)\sigma_3}\ ,\qquad
0\leqslant\omega\leqslant\pi/2 \ ,\quad -\pi\leqslant \phi_{1,2}<\pi\ .
\ee
The background metric and antisymmetric tensor of the corresponding $\s$-model are
\be
\begin{split}
\label{backwzw}
& \dd s^2= k\left(\dd\omega^2 + \cos^2\omega \dd\phi_1^2 + \sin^2\omega
\dd\phi_2^2\right)\ ,
\\
& B =k\sin^2\omega\ \dd\phi_1\wedge \dd\phi_2
\end{split}
\ee
and the dilaton is constant. Consider zooming in near a point. Specifically, take the limit
\begin{align}
\omega=\frac{\rho}{\sqrt{k}}\ ,\qquad \phi_1=\frac{z}{\sqrt{k}}\ ,\qquad \phi_2=\phi\ ,
\qquad k\to\infty\ ,
\end{align}
where because of the limit the range of variables is
$0\leqslant \rho<\infty$, $-\pi\leqslant\phi<\pi$ and $-\infty <z< \infty$.
The limiting background is
\begin{align}
\label{eq:flatspace}
\dd s^2=\dd\rho^2 + \rho^2\dd\phi^2 + \dd z^2\ ,\qq B=0 \ .
\end{align}
i.e. free motion in $\mathbb{R}^3$, which is expected given that we zoomed covariantly with respect to $SU(2)$ -
 it is just the tangent space at a point.

\no
We would like to see how this limit can be taken at the level of the spectrum of the
theory and the associated partition function.

\subsection{The spectrum}\label{subsec:The wzw spectrum}

We now turn to the quantization of the theory.
It is relevant to mention that the quantization of a given classical theory is not unique.
Therefore we should be clear about which quantum theory we are considering: it is the quantization which preserves
all the chiral symmetries of the WZW Lagrangian corresponding to the background \eqref{backwzw}, leading to conformal invariance.
The spectrum of the theory fits into representations of the infinite-dimensional affine Lie algebra $\hat{\mathfrak{su}}(2)_k$.
In fact, given consistency requirements such as unitarity and a bounded spectrum,
only the `integrable highest weight' representations appear, which for brevity we will call simply `integrable'.
For further details we refer the reader to \cite{DiFrancesco:1997nk}.
The Hilbert space decomposes in terms of $\hat{\mathfrak{su}}(2)_k$ representations as a diagonal sum
\be\label{eq:H decomposition}
\mathcal{H}_\mt{WZW} = \bigoplus_{\lambda}R_{\lambda} \otimes R_{\lambda}\ ,
\ee
with the sum over all representations at level $k$, and every representation appears exactly once.
This means that all primary fields have equal holomorphic and antiholomorphic conformal dimension.

An integrable representation is specified by an integer $0\leqslant \lambda\leqslant k$.
This is just twice the usual spin quantum number $\lambda = 2j$ ($j=0,1/2,1,\dots, k/2$), more common in the physics literature.
The Virasoro central charge and holomorphic conformal dimensions are given by
\begin{align}
c=\frac{3k}{k+2} \qquad \mathrm{and} \qquad h_{\lambda}^{(k)}
= \frac{\lambda (\lambda + 2)}{4(k+2)}\ .
\end{align}
From \eqref{eq:H decomposition} the partition function is given by a bilinear combination of the characters $\chi^{(k)}_{\lambda}(\tau)$ of the integrable representations of $\hat{\mathfrak{su}}(2)_k$
\be
\label{eq:wzw diagonal part function}
\mathcal{Z}_\mt{WZW}=\sum_{\lambda =0}^k |\chi^{(k)}_{\lambda}(\tau)|^2 \ .
\ee
Here $\tau$ is the modular parameter of the torus, such that the identifications are $z\sim z +1\sim z+\tau$. We also define $q\equiv \mathrm{e}^{2\pi \mathrm{i} \tau}$, and in an abuse of notation refer for example to $\chi^{(k)}_{\lambda}(\tau)$ also as $\chi^{(k)}_{\lambda}(q)$ when it suits us.
The partition function is modular invariant since the characters transform into one another under the action of the modular group\footnote{The modular group $PSL(2,\mathds{Z})$ is generated by the two transformations
\be\label{eq:S,T maps}
\mathcal{T} \colon \tau \mapsto \tau+1, \quad \mathcal{S} \colon \tau \mapsto -\frac{1}{\tau} ~,
\ee
which act on the upper half plane $\mathds{H}=\{\tau\in\mathds{C}\mid \Img\tau>0 \}$.}
\bea\label{eq:character transforms}
\chi^{(k)}_\l(\tau+1)=\sum_{\l'}\mathcal{T}_{\l\l'}\chi^{(k)}_{\l'}(\tau)\ ,
\\
\displaystyle\chi^{(k)}_\l\(-\frac{1}{\tau}\)=\sum_{\l'}\mathcal{S}_{\l\l'}\chi^{(k)}_{\l'}(\tau)\ .
\eear
\ee
The explicit expression for the $\hat{\mathfrak{su}}(2)_k$ characters can be obtained from the Kac-Weyl formula
and reads
\begin{align}
\label{su2characteroriginal}
\chi^{(k)}_{\lambda}(q) = \frac{q^{(\lambda +1)^2/4(k+2)}}{\eta (q)^3}
\sum_{n=-\infty}^{\infty} (\lambda +1+2n(k+2))\, q^{ n[\lambda + 1 + (k+2)n ] } \ ,
\end{align}
where $\displaystyle \eta (q)\equiv q^{\frac{1}{24}}\prod_{n=1}^{\infty}(1-q^n)$ is the Dedekind eta-function. See Appendix \ref{factorized_su2} for more information on the affine characters.

\subsection{Large level limit}
We now want to find which states we have to keep in the quantum theory in order to recover flat space,
i.e. free bosons, when we take the large-$k$ limit of the $\hat{\mathfrak{su}}(2)_k$ WZW model.
In addition, we would like to see if there is a truncation of the above character which still gives a modular invariant partition function
as $k\rightarrow\infty$. Indeed, in this limit, only the $n=0$ term in the sum in \eqn{su2characteroriginal} survives
\begin{align}
\label{largeksu2}
\chi^{(k)}_{\lambda}(q)\big|_{k\to\infty}\, =\, \frac{q^{(\lambda +1)^2/4(k+2)}}{\eta (q)^3}(\lambda +1)\ .
\end{align}
There are three distinct natural scalings of $\l$ as $k\to \infty$. We consider each in turn, substituting \eqref{largeksu2} into \eqref{eq:wzw diagonal part function} to find the partition function:
\begin{itemize}
\item Take all $\lambda={\cal O}(1)$, i.e. kept finite as we scale $k\rightarrow\infty$. Then all the primary fields have dimension zero, and the power in the numerator
in \eqref{largeksu2} vanishes. We get $\chi^{(k)}_{\lambda}(q)=(\lambda+1)/\eta (q)^3$
and the contribution of these states for the diagonal partition function is
\begin{align}
\label{finitepartition}
\mathcal{Z}_\mt{WZW,\ 1}=\frac{\sum_{\lambda =0}^{\lambda_{\rm max}}(\lambda +1)^2}{|\eta (q)|^6}
\sim 
\frac{1}{|\eta (q)|^6}\ ,
\end{align}
where by assumption $\lambda_{\rm max}$ is some order one constant.
That is to say, in taking this limit the $\mathcal{O}(1)$ spins contribute $\mathcal{O}(1)$ in total to the partition function.
The result, \eqref{finitepartition}, is not modular invariant and as such a corresponding truncated theory is not consistent.

\item
Now let $\lambda ={\cal O}(\sqrt{k})$, which can be  parametrized as $\lambda = \gamma\sqrt{k}$, with $\g > 0$.
The weight of the primary field becomes $h_{\lambda}^{(k)} \rightarrow \gamma^2/4$.
 We can work out the contribution of these states to the
  partition function by replacing the sum with an integral
\be
\begin{split}
\mathcal{Z}_{\mt{WZW,}\sqrt{k}} &= {1\ov |\eta (q)|^6}
\int_0^\infty \dd\lambda  \lambda^2  |q|^{\lambda^2\ov 2k}
=
\frac{k^{3/2}}{|\eta (q)|^6}\int_0^{\infty} \mathrm{d}\g \g^2
\mathrm{e}^{-\pi \mathrm{Im}\tau \g^2}
\\
&=k^{3/2}\left( \frac{1}{\sqrt{\mathrm{Im} \tau}}\frac{1}{|\eta (q)|^2} \right)^3
\\
&=k^{3/2}\left( \mathcal{Z}_{\mathrm{free\, boson\, on}\, \mathbb{R}} \right)^3\ .
\end{split}
\ee
This is the partition function for the background \eqref{eq:flatspace} since in Cartesian coordinates
it corresponds to the a $\s$-model action for three free bosons. As such each factor is modular invariant separately.
The $k^{3/2}$ scaling factor is proportional to the volume of the group manifold.

\item Finally, let $\lambda ={\cal O}(k)$. All the dimensions of the primaries become very large -
of order $ k$. Thus the characters exponentially vanish and so does the entire partition function.
\end{itemize}
The main point we want to make in this section is the following: it is only a \textit{subset} of primary states we need to consider in order to find flat space as a limit of the WZW model. This is the subset of weights scaling with $\lambda\sim\sqrt{k}$. The situation will be similar when we consider coset models, except that the scaling of the weights will be more subtle.

\subsection{Modular \texorpdfstring{$\mathcal{S}$}{S}-matrix}
\label{sec: wzwmod}
We have seen that the subset of states in the $\hat{\mathfrak{su}}(2)_k$ modules whose primaries scale like $\sqrt{k}$ can be picked out, and form on their own a modular invariant partition function.
We would like to arrive at the same conclusion by examining directly how modular transformations act on the $\hat{\mathfrak{su}}(2)_k$ characters.
We will focus on the transformation $\displaystyle \mathcal{S}\colon \tau\mapsto -\frac{1}{\tau}$,
in \eqref{eq:character transforms} implemented by the modular $\mathcal{S}$-matrix with elements
\be
\label{eq:S matrix su2}
\mathcal{S}^{(k)}_{\lambda\lambda'}=\sqrt{\frac{2}{k+2}}\sin \left( \pi\frac{(\lambda+1)(\lambda'+1)}{k+2} \right)\ .
\ee
This $\mathcal{S}$-matrix is real and symmetric, hence unitary, and squares  to unity.
We would like to study how weights scaling with different powers of $k$ are related to each other by the modular $\mathcal{S}$-transformation. First note that at large $k$ we have approximately,
after dropping an overall numerical factor, that
\begin{align}
\label{larges}
\mathcal{S}^{(k)}_{\lambda\lambda'} \sim \frac{1}{\sqrt{k}}\sin \left( \pi\frac{\lambda\lambda'}{k} \right)\ .
\end{align}

We consider again the diagonal partition function \eqref{eq:wzw diagonal part function}, now in terms of a power counting at large $k$. There are three sources of scaling we must taking into account: the scaling of the characters, the number of characters in that regime of $\lambda$, and the scaling of the relevant matrix element of $\mathcal{S}^{(k)}$. In a schematic notation, we denote these by $\chi_{\lambda}$, $n_{\lambda}$ and $\mathcal{S}_{\lambda,\lambda'}$ respectively.

Let us divide the weights into the three regimes $\lambda\sim 1$, $\sqrt{k}$, $k$, as
before.  Obviously, since a representation is characterized uniquely by $\l$,
the number of $\lambda\sim 1$ weights is of order one, so we write $n_1\sim 1$ and similarly $n_{\sqrt{k}}\sim \sqrt{k}$ and $n_k\sim k$. Then we look at the character formula \eqref{su2characteroriginal}. The characters are order one for $\lambda\sim 1$, order $\sqrt{k}$ for $\lambda\sim\sqrt{k}$, but for $\lambda\sim k$ all of the weights are sent to infinity and the
corresponding character goes to zero, exponentially for large $k$.
We may tabulate this information as
\begin{align}
\begin{array}{ccccc}
\lambda&&n_{\sqrt{\lambda}}&&\chi_{\lambda}\\
\hline
1&&1&&1\\
\sqrt{k}&&\sqrt{k}&&\sqrt{k}\\
k&&k&&\mathrm{e}^{-k}
\end{array}
\end{align}

Using this we can write the diagonal partition schematically as
\be
\label{schematicz}
\mathcal{Z}_\mt{WZW} = \sum_{\lambda}|\chi_{\lambda}|^2\, \sim\, n_1\chi_{1}^2 + n_{\sqrt{k}}\chi_{\sqrt{k}}^2 + n_k\chi_k^2\,\sim\,
1+k^{3/2}+\mathrm{e}^{-k}\ .
\ee
This shows the leading large-$k$ contributions from each weight regime. We see that the dominant contribution is the middle term coming from $\mathcal{O}(\sqrt{k})$ weights. The large-$k$ expansion of the partition function is $\mathcal{Z}_{\rm WZW}=k^{3/2}\mathcal{Z}_0+\cdots$, and $\mathcal{Z}_0$ can be computed using only these weights.\footnote{We can make an analogy with the canonical ensemble in statistical mechanics, where $k$ plays the role of the particle number. In that case there is a competition between entropy and energy. The entropy of the $\mathcal{O}(1)$ states is too low, and the energy of the $\mathcal{O}(k)$ states is too high. In the thermodynamic limit, the dominant contribution to the free energy is from states where the two effects balance, namely for $\lambda={\cal O}(\sqrt{k})$.} Now let's do a modular $\mathcal{S}$ transformation on \eqref{schematicz} and see how the different contributions map into each other. We keep track of the various contributions to the dominant $k^{3/2}$ term
\begin{multline}\label{schematicz_prime}
  \mathcal{Z}_\mt{WZW}\left( -\frac{1}{\tau} \right)=\sum_{\lambda} \sum_{\lambda'} |\mathcal{S}_{\lambda\lambda'}\chi_{\lambda'}|^2
\sim n_{\sqrt{k}} \left( n_1\chi_1 \mathcal{S}_{\sqrt{k}, 1} + n_{\sqrt{k}}\chi_{\sqrt{k}}\mathcal{S}_{\sqrt{k},\sqrt{k}} + n_k\chi_k \mathcal{S}_{\sqrt{k},k} \right)^2 \\
  \sim\sqrt{k}\left(\frac{1}{k}+\sqrt{k}+k\cdot \mathrm{e}^{-k}\right)^2\sim k^{3/2}.
\end{multline}
The overall scaling behavior \eqref{schematicz} (i.e. $k^{3/2}$) is recovered by only considering the middle term in \eqref{schematicz_prime}.  Even though modular transformations mix large weights with small weights, as far the as the leading term in the partition function is concerned the only important weights are $\mathcal{O}(\sqrt{k})$. The other weights can be consistently ignored, even when considering modular invariance, since they simply do not contribute to leading order.

The special property of $\mathcal{O}(\sqrt{k})$ weights is that their primaries have finite non-vanishing conformal dimension in the $k\rightarrow\infty$ limit. Likely this is true for general
WZW theories: their modular behavior at large level should be dominated by the states which have finite non-zero dimension. Only these will be closed under modular transformations.
In the coset theory we will encounter a similar situation.

\subsection{Fusion rules}
\label{subsec:fusion rules}
Lastly we consider briefly the fusion rules of the theory. The consistency of the OPE with the spectrum \eqref{eq:H decomposition} implies that the product of two representations of $\hat{\mathfrak{su}}(2)_k$ can be decomposed as
\be\label{eq:fusion rules wzw}
R_\l \otimes R_\mu = \sum_{\nu=0}^k \mathcal{N}_{\l\mu}^{(k)\phantom{\l}\nu}R_\nu\ ,
\ee
where the fusion coefficients $\mathcal{N}_{\l\mu}^{(k)\phantom{\l}\nu}$ count the
number of times a given representation appears in the decomposition.
Consistency of the CFT on the torus means they have an expression in terms of the modular matrix given by the
Verlinde formula \cite{Verlinde:1988sn}
\begin{align}
\label{verlinde}
\mathcal{N}_{\l\mu}^{(k)\ \nu}=\sum_{\rho} \frac{\mathcal{S}\mathcal{}_{\lambda\rho}^{(k)}\mathcal{S}_{\mu\rho}^{(k)}\mathcal{S}_{\nu\rho}^{(k)}}{\mathcal{S}_{0\rho}^{(k)}},
\end{align}
where $0$ denotes the vacuum representation. It follows from the Kac--Walton formula relating tensor-product to fusion coefficients \cite[Sect. 16.2]{DiFrancesco:1997nk} that the fusion coefficients reduce to tensor-product coefficients in the limit $k\to\infty$
\be\label{eq:lim of N coef}
\lim_{k\to\infty}\mathcal{N}_{\l\mu}^{(k)\ \nu}=\mathcal{N}_{\l\mu}^{\phantom{\l\mu}\nu}.
\ee
In other words, the fusion coefficients are some sort of truncated tensor-product coefficients $\mathcal{N}_{\l\mu}^{\phantom{\l\mu}\nu}$. We would like to see whether there is a consistent truncation of this statement to only the $\lambda\sim\mathcal{O}(\sqrt{k})$ states - namely, whether a theory containing only these states can reproduce \eqref{eq:lim of N coef}. So first we take all the states including the intermediate states in the sum $(\lambda,\mu,\nu,\rho)\equiv(\sqrt{k}\ \tilde{\lambda},\sqrt{k}\ \tilde{\mu},\sqrt{k}\ \tilde{\nu},\sqrt{k}\, x)$ to have this scaling. We turn the sum over $\rho$ into an integral over $x$ to obtain that
\begin{align}
\label{su2fusionint}
\mathcal{N}_{\tilde{\l}\tilde{\mu}}^{(k)\ \tilde{\nu}}|_{k\rightarrow\infty}= \frac{2}{\pi}\int_0^{\infty}\frac{\dd x}{x} \sin (\pi\tilde{\lambda} x)\sin (\pi\tilde{\mu} x)\sin (\pi\tilde{\nu} x)\ .
\end{align}
Assuming that none of $\tilde{\lambda}\pm\tilde{\mu}\pm\tilde{\nu}$ vanishes,\footnote{Strictly speaking there are subtleties in taking this sort of limit, but our goal is simply to show a truncation to a certain subset of states, so we will skim over these.} the integral converges and we get
\begin{align}
\mathcal{N}_{\tilde{\l}\tilde{\mu}}^{(k)\phantom{\l}\tilde{\nu}}\Big|_{k\rightarrow\infty}=\begin{cases}
1,&|\tilde{\lambda}-\tilde{\mu}|<\tilde{\nu}<\tilde{\lambda}+\tilde{\mu}
\\
0,&\mathrm{otherwise}\ .
\end{cases}
\end{align}
This makes sense in terms of $SU(2)$ representations: it is the selection rule for adding two spins $\lambda$ and $\mu$ to get resultant spin $\nu$. It also confirms that \eqref{eq:lim of N coef} is valid within our truncation.
Using \eqref{su2fusionint} we can check that no $\mathcal{O}(k)$ mode appears in the decomposition of two $\mathcal{O}(\sqrt{k})$ modes: for large $\tilde{\nu}$ the integrand becomes rapidly oscillatory and the integral vanishes.


\section{Coset CFT}
\label{sec:Coset th}

Now we consider the coset \eqref{eq:coset su2 diagonal}. In general, cosets are more complicated than WZW models to study, but in this simple case both the background fields and the modular invariant partition function are known.

\subsection{The background}

The background fields for this coset were first worked out in \cite{Bardacki:1990wj} and in \cite{Bars:1991pt,Bars:1992ti} (for the
equal level case and Minkowskian signature generalizations) using the gauged WZW action. Here we present the results of the derivation of \cite{Polychronakos:2010fg}
in which the parametrization used resulted in expressions more convenient for our purposes.

We start with the standard parametrization of the two $SU(2)$ group elements
\be
g_1 =
\left(
  \begin{array}{cc}
    \a_0 + \ii \a_3 & \a_2 + \ii \a_1 \\
    -\a_2+\ii \a_1 & \a_0 - \ii \a_3 \\
  \end{array}
\right)\ ,\qq g_2 =
\left(
  \begin{array}{cc}
    \b_0 + \ii \b_3 & \b_2 + \ii \b_1 \\
    -\b_2+\ii \b_1 & \b_0 - \ii \b_3 \\
  \end{array}
\right)\ ,
\label{e-2-8a}
\ee
where from unitarity
\be
\a_0^2 + \mathbold{\a}^2 = 1\ , \qq \b_0^2 + \mathbold{\b}^2 = 1\ .
\label{e-2-8b}
\ee
The gauging procedure from which the $\s$-model background fields are derived implies that resulting three dimensional
space will depend only on the rotationally invariant combinations of the two vectors $\mathbold{\alpha}\equiv (\alpha_1,\alpha_2,\alpha_3)$ and $\mathbold{\beta}\equiv (\beta_1,\beta_2,\beta_3)$.
These are their lengths and their inner product. We will use equivalently, $\a_0$, $\b_0$ and
$\g\equiv \sqrt{\mathbold{\pm \alpha\cdot\beta}}$ (depending if the inner product is positive or negative. The results below
do not depend on this sign).

\no
It is convenient to introduce the ratio of levels
\be
r \equiv {k \ov \ell} \ .
\ee
Then, the background metric is given by
\begin{multline}\label{cosetmetric}
   \dd s^2 = {k + \ell \ov (1-\a_0^2 ) (1-\b_0^2 ) - \g^2}
\bigl( \D_{\a\a} \dd\a_0^2 + \D_{\b\b} \dd\b_0^2 + \D_{\g\g} \dd\g^2 \\
   + 2 \D_{\a\b} \dd\a_0 \dd\b_0 + 2 \D_{\a\g} \dd\a_0 \dd\g + 2 \D_{\b\g} \dd\b_0 \dd\g \bigr) \quad ,
 \end{multline}
the antisymmetric tensor is zero, and the dilaton reads
\be
\label{cosetfields}
\mathrm{e}^{-2 \Phi} =   (1-\a_0^2 ) (1-\b_0^2 ) - \g^2 \ .
\ee
The definitions of the various functions are
\ba
&&\D_{\a\a} = {(1+r)^2 -r(2+r) \b_0^2 \ov r(1+r)^2}\ ,
\qq \D_{\b\b} = {(1+r^{-1})^2 -r^{-1}(2+r^{-1}) \a_0^2\ov r^{-1} (1+r^{-1})^2}\ ,
\nonumber\\
&&\D_{\g\g} = {1 \ov 2 + r +r^{-1}}\ ,
\qq \phantom{xx}
\D_{\a\b} = \g + {\a_0 \b_0 \ov 2+r+r^{-1}} \ ,\\
&&\D_{\a\g} = -{ \b_0 \ov (1+r)^2}\ ,
\qq \phantom{xx} \D_{\b\g} = -{ \a_0 \ov (1+r^{-1})^2} \ .
\nonumber
\ea
The range of variables is
\be
|\a_0|,|\b_0|\leqslant 1\ ,\qq \g\leqslant \sqrt{(1-\a_0^2 ) (1-\b_0^2 )}\ .
\ee
The background is manifestly invariant under the interchange of $\a_0$ and $\b_0$ and a simultaneous inversion
of the parameter $r$. This symmetry simply interchanges the two $SU(2)$ factors in the coset.

\subsubsection{High spin limit and the effective geometries}

We are interesting in taking one of the levels, say $\ell$, infinitely large. In accordance with
our discussion for the group case this should be done in such a way that the eigenvalues of the
Hamiltonian, i.e. of the zero modes of the Virasoro algebra generators $L_0+\bar L_0$,
remain finite. Using the $\s$-model background these eigenvalues are computed
using the Laplace operator corresponding to  the metric \eqn{cosetmetric}
and the dilaton \eqn{cosetfields}. This was done in \cite{Polychronakos:2010fg,Polychronakos:2010hd}
and involved new group theoretical methods since \eqn{cosetmetric}
lacks isometries. A typical eigenstate is labelled by the non-negative integers $\m,\l,\n$.
The corresponding eigenvalues are in general non-degenerate and given by
\be
E^\n_{\m,\l} = {\m(\m+2)\ov 4\ell} + {\l(\l+2)\ov 4k} - {\n(\n+2)\ov 4(k+\ell)}\ .
\label{enrjjj}
\ee
Given a pair of values for ($\m,\l$) then $\n=|\m-\l|, \dots,  \m+\l$.
The above expression is
nothing but the conformal weight of the corresponding coset CFT primary field in the
semi-classical limit of large levels in which a shift in the denominators of all expressions by two (the
dual Coxeter number for $SU(2)$) has been omitted.

\no
When sending $\ell\to \infty$, a non-trivial result is achieved provided one of the spins
becomes large, whereas the other one is kept finite. Specifically, let \cite{Polychronakos:2010fg,Polychronakos:2010hd}
\be
\label{geometrylimit}
\nu = \mu + n\ ,\qq \mu = \frac{\delta}{k}\ell \ ,
\ee
where $n$ is a integer and $\d$ a positive real parameter.
Then
\be E_{\l,n,\d}\equiv \lim_{\ell\to
\infty} E^\n_{\m,\l}= {\l(\l+2)\ov 4 k} +  {\d - 2 n\ov 4 k}\ \d\ .
\label{eiginfi}
\ee
Note that in this limit the first and the third terms in \eqn{enrjjj} become separately infinite, but their sum remains finite as is
seen above.
The above limit corresponds to zooming into the metric to find the
geometry that can support this infinite level/spin regime in a consistent manner.
This can be done in two different ways resulting in two different gravitational backgrounds \cite{Polychronakos:2010fg}.

 \no
 {\bf Limit 1:} Consider focusing around $\a_0=0$.
Accordingly, we scale this coordinate and take the limit
\be
\a_0 \equiv r z\ ,\qq r={k\ov \ell}\to 0\ ,
\ee
where $z$ is the new uncompactified coordinate. After performing the
transformation
\be
\g \equiv \sin\th \cos(\phi + z) \ ,\qq \b_0 \equiv \sin\th \sin(\phi + z)\ ,
\label{jdperi}
\ee
we obtain that
\be
\dd s^2 =k( \dd z^2 + \dd \th^2 + \tan^2\th\ \dd\phi^2)\ ,\qq \mathrm{e}^{-2\Phi} = \cos^2\th\ .
\label{hgeq}
\ee
This is locally the background for the exact CFT $\hat{\mathfrak{su}}(2)_{k}/\hat{\mathfrak{u}}(1) \times \mathbb{R}$. However globally there is a twisting, as can be  seen from the coordinate change \eqref{jdperi}. There we see it is not $\phi$ but $\phi +z$ which is the cyclic coordinate: the coordinate ranges in \eqref{hgeq} are in fact $0\leqslant\phi +z<2\pi$, $-\infty<z<\infty$.

The spectrum corresponding to \eqn{hgeq} should be of the form \eqn{eiginfi}. To see this, note that
the eigenvalues of the Laplace operator for \eqn{hgeq} are given by
$ \displaystyle {1\ov 4k}\big(\l(\l+2)-a^2+b^2\big)$ as expected for a coset CFT space
supplemented by a free extra coordinate.
Because of translation invariance in $\phi $ and $z$, this corresponds to an eigenfunction proportional to $\mathrm{e}^{\mathrm{i}a\phi}\mathrm{e}^{\mathrm{i}b z}$.
This factor can be rewritten as $\mathrm{e}^{\mathrm{i}a(\phi+z)}\mathrm{e}^{\mathrm{i}(b-a) z}$. Since $\phi+z$ is periodic
then $a= n$, with $n\in \mathbb{Z}$. On the other hand $a-b= \d$, with $\d\in \mathbb{R}$, so that
$b=n-\d$. Hence $-a^2+b^2= \d(\d-2 n)$, and we reproduce \eqn{eiginfi}.

 \no
 {\bf Limit 2:}
Consider focusing around
$\a_0 =1$ and $\g =0$, by performing first the coordinate transformation
\be
\label{limit2}
\a_0^2 = 1 - r^2 \left[(x_1+\psi )^2 + x_3^2 \right]\ ,
\qq \g = r (x_1+\psi) \cos\psi\ , \qq \b_0 = \sin \psi\ ,
\ee
followed by the limit $r\to 0$. Then the new variables $x_1$ and $x_3$, that we will
use instead of $\a_0$ and $\g$, become uncompactified.
In this limit we obtain for the metric and dilaton the expressions
\be
\dd s^2 =k\left(\dd\psi^2 + {\cos^2 \psi\ov x_3^2}\dd x_1^2
+ {\big[ x_3 \dd x_3 + (\sin\psi\cos\psi +  x_1+\psi) \dd x_1\big]^2\ov x_3^2\cos^2\psi}\right)
\label{non1}
\ee
and
\be
\mathrm{e}^{-2 \Phi} = x_3^2 \cos^2 \psi\ .
\label{non2}
\ee
Remarkably, the Laplace operator on scalars for this background also has eigenvalues given by \eqn{eiginfi} \cite{Polychronakos:2010fg,Polychronakos:2010hd}. This follows from the fact that both backgrounds come from the same limit in the quantum numbers, but is completely non-trivial from the geometrical point of view since the metrics \eqn{hgeq} and \eqref{non1} differ drastically.

The limit \eqref{limit2} is a particular case of a more general limiting procedure mentioned in the introduction
\cite{Sfetsos:1994vz} in which the action for the non-Abelian T-dual of the WZW model with respect to the vector symmetry $g\mapsto h\, g\, h^{-1}$ coincides with the $\ell\rightarrow\infty$ limit of the gauged WZW action if we simultaneously zoom in near the identity $
g_2=\mathds{1}+\mathrm{i}\ \frac{v}{\ell}$, with  $\ell\rightarrow\infty$.

The above limits suggest that two different backgrounds, one of which is the non-Abelian T-dual of the $\hat{\mathfrak{su}}(2)$ WZW model,
describe in fact the same consistent sector of representations with very high values for the Casimir operators.
Next we will take the same limit at the level of the modular invariant partition function of the coset theory.
%
%
%
%
%
\subsection{The quantum theory}
\label{cosetspectrum}
The $\hat{\mathfrak{su}}(2)_k\oplus \hat{\mathfrak{su}}(2)_{\ell} / \hat{\mathfrak{su}}(2)_{k+\ell}$ coset theory is a quantization of \eqref{cosetmetric}, \eqref{cosetfields} that preserves its group theoretical structure. The corresponding CFT coset construction gives an exact conformal theory whose chiral data (conformal weights, fusion rules, etc.) can be expressed entirely in terms of the chiral data for the $\hat{\mathfrak{su}}(2)_k$ and $\hat{\mathfrak{su}}(2)_{\ell}$ WZW models.

A coset field is labelled by three highest weights $\{\lambda , \mu ; \nu\}$ of $\hat{\mathfrak{su}}(2)_k$, $\hat{\mathfrak{su}}(2)_{\ell}$ and $\hat{\mathfrak{su}}(2)_{k+\ell}$ respectively. We have that $0\leqslant \lambda\leqslant k$, $0\leqslant \mu\leqslant \ell$, and $0\leqslant \nu\leqslant k+\ell$, but this overcounts the coset fields since we must make the identification (corresponding to simultaneous automorphisms of the algebras)
\begin{align}
\label{identification}
\{\lambda , \mu ; \nu\}\sim \{k-\lambda , \ell -\mu ; k+\ell-\nu\}\ ,
\end{align}
and we will also find the selection rule
\begin{align}
\label{selectionrule}
\lambda +\mu - \nu\in 2\mathbb{Z}\ .
\end{align}

In order to extract the coset Hilbert space we must decompose each representation of $\hat{\mathfrak{su}}(2)_k \oplus \hat{\mathfrak{su}}(2)_{\ell}$ into a direct sum of representations of $\hat{\mathfrak{su}}(2)_{k+\ell}$
\be\label{eq:rep decomposition}
R_{\l}^{\hat{\mathfrak{su}}(2)_k} \otimes R_{\mu}^{\hat{\mathfrak{su}}(2)_{\ell}} = \bigoplus_{\nu=0}^{k+\ell} \Hc_{\{\l,\mu;\nu\}} \otimes R_\nu^{\hat{\mathfrak{su}}(2)_{k+\ell}}\  .
\ee
The Hilbert space of the coset theory is a diagonal sum of coset modules
\be\label{eq:coset H space}
\mathcal{H}_{\rm coset} = \bigoplus_{\l,\mu,\nu}\Hc_{\{\l,\mu;\nu\}}\otimes\overline{\Hc}_{\{\l,\mu;\nu\}}\ ,
\ee
where the two tensor factors are representations of the left and right moving coset CFTs respectively, and the sum is subject to \eqref{identification} and \eqref{selectionrule}.

The characters of the coset are obtained from the character identity corresponding to \eqref{eq:rep decomposition}
\be\label{eq:char decomposition id}
\chi^{(k)}_{\l}(z,\tau)\chi^{(\ell)}_{\mu}(z,\tau) = \sum_{\nu=0}^{k+\ell}\Xi_{\{\l,\mu;\nu\}}(\tau)\chi^{(k+\ell)}_{\nu}(z,\tau)\ .
\ee
The functions $\Xi_{\{\l,\mu;\nu\}}(\tau)$ are called the coset characters of the coset field $\{ \l,\mu;\nu \}$. Under the modular $\mathcal{S}$ transformation they transform into one another as
\be\label{eq:T coset}
\Xi_{\{\l,\mu;\nu\}}\(-\frac{1}{\tau}\) = \sum_{\l',\mu',\nu'}\mathcal{S}^{(k)}_{\l\l'}\mathcal{S}^{(\ell)}_{\mu\mu'}\mathcal{S}^{(k+\ell)}_{\nu\nu'}\Xi_{\{\l',\mu';\nu'\}}(\tau)\ .
\ee
The modular invariant partition function is given by \cite{Gawedzki:1988nj}
\be\label{eq:coset partition function}
\mathcal{Z}_\mathrm{coset}=\sum_{\l,\mu,\nu} |\Xi_{\{\l , \mu ; \nu\}}(\tau)|^2\ ,
\ee
subject to \eqref{identification} and \eqref{selectionrule}.

\subsection{The spectrum}
We are interested in taking the $\ell\rightarrow\infty$ limit of the coset theory. The analysis is greatly simplified by using a form of the $\mathfrak{su}(2)_k$ characters factorized in terms of the `parafermion' theory and a $\hat{\mathfrak{u}}(1)_k$ model (single compactified boson), material which is presented in appendix \ref{factorized_su2}. This corresponds to considering the group manifold as a Hopf fibration of $U(1)$ over $SU(2)/U(1)$, i.e. $ S^1$ over $S^2$. The boson $S^1$ is the Hopf fibre, and is the maximal torus of $SU(2)$. The coset construction leaves intact the parafermion characters, while modifying the $\hat{\mathfrak{u}}(1)_k$ factors in a way we will explore in this section.

In appendix \ref{characterderivation} we derive a simple expression for the coset character in the limit $\ell\rightarrow\infty$, in terms of parafermions
\be
\label{cosetcharacter}
\begin{split}
&
\Xi_{\{\lambda , \mu ; \nu\}}= c^{\lambda}_{\nu-\mu} (q)\; q^{\beta_{\mu,\nu}}\, -\, c^{\lambda}_{\nu+\mu -2} (q)\; q^{\beta_{\mu,2-\nu}}\ ,
\\
&
\beta_{\mu,\nu}\equiv \frac{[(k+\ell)(\mu + 1)-\ell(\nu + 1)]^2}{4k\, \ell(k+\ell)} \ .
\end{split}
\ee
The second term projects out a subset of states of the $\hat{\mathfrak{u}}(1)_k$ factor of the theory, the coset analogue of singular states in a Virasoro module.
Notice that in \eqref{stringnormalization} we defined the functions $c_m^{\lambda}(q)$ to vanish for $\lambda - m$ odd, which guarantees that the selection rule \eqref{selectionrule} is satisfied. In order to deal with the field identification \eqref{identification} so as to cover each independent field only once, we can for example restrict $\mu-\nu\leqslant \frac{k}{2}$. However, for generic weights \eqref{identification} acts transitively, and we simply divide the partition function by 2 - so there is no non-trivial constraint on $\{\lambda,\mu,\nu\}$ in the above expressions.

The primary state of a coset module is the state of lowest conformal dimension. We can read this off as the lowest power of $q$ appearing in the expansion of \eqref{cosetcharacter}.  There are two contributions: one from the exponent $\beta_{\mu,\nu}$ and one from the normalization \eqref{stringnormalization} of the string functions
\begin{align}
h_{\{\lambda , \mu ; \nu\}}\, =\, \frac{[(k+\ell)(\mu + 1)-\ell(\nu + 1)]^2}{4k\, \ell(k+\ell)}\, +\, h^{\lambda}_{\mu-\nu} (\mathrm{parafermion})\ ,
\end{align}
where the definition of $h^{\lambda}_{\mu} (\mathrm{parafermion})$ in \eqn{stringnormalization} has been used.
Comparing this with the `na\"{i}ve'  weight $h_{\lambda}^{(k)}+h_{\mu}^{(\ell)}-h_{\nu}^{(k+\ell)}$  there is an integer shift in the dimension of the primary \cite{Dunbar:1992gh}. Indeed,
because the parafermions have the shift symmetry $c^{\lambda}_m=c^{\lambda}_{m+2k}$ it is always convenient to write $\nu-\mu$ modulo $2k$ in the following way
\begin{align}
\label{mod2k}
\nu-\mu=2k\,\alpha +\beta \ ,\quad \alpha , \, \beta\in\mathbb{Z}\ ,\quad -k+1\leqslant\beta\leqslant k\ ,
\end{align}
so that $c^{\lambda}_{\mu-\nu} = c^{\lambda}_{\beta}$. Using this we can write the integer shift $s_{\lambda,\mu,\nu}$ as
\be
\label{dimension}
\begin{split}
& h_{\{\lambda , \mu ; \nu\}}= h_{\lambda}^{(k)}+h_{\mu}^{(\ell)}-h_{\nu}^{(k+\ell)} + s_{\lambda,\mu,\nu}\ ,
 \\
& s_{\lambda,\mu,\nu}= \alpha\beta + k\alpha^2 + \left(\frac{|\beta|-\lambda}{2}\right)\theta (|\beta|-\lambda)\ ,
\end{split}
\ee
where $\theta (x)$ is the Heaviside function
\begin{align}
\theta (x)\, =\, \begin{cases}0&x<0\ ,\\1&x\geqslant 0\ .
\end{cases}
\end{align}

\subsection{Limit of the partition function}
We now want to establish the main result of this paper by taking the $\ell\rightarrow\infty$ limit in the partition function \eqref{eq:coset partition function}. As before, the question is whether all coset primary states are relevant in this limit. In particular, is there a subset of states closed under modular transformations? The results in the case of the WZW model suggest that the modular invariant limit will correspond to scaling weights with the highest power of $\ell$ such that conformal dimensions of the primaries remain finite. Indeed, this will be precisely the same limit as \eqref{geometrylimit} in the geometry.

\subsubsection{Linear scaling}
\label{sec:linear}
As in \eqn{geometrylimit}, in this limit we take both $\mu$ and $\nu$ to scale linearly with $\ell$, but with their difference $\mu-\nu$ kept finite:
\begin{align}
\label{linearscaling}
\mu = \frac{\delta}{k}\, \ell\ , \qquad \nu = \mu + n\ ,\qquad \ell\rightarrow\infty\ ,
\end{align}
where $0\leqslant\delta \leqslant k$ becomes a continuous variable in the limit, and $-\infty < n <\infty$.\footnote{
Notice that in non-affine group theory we have $ |n|\leqslant \lambda$. In the affine case there is no restriction on the range of $n$. But the conformal dimensions get very large for $|n|\gg\lambda$ and so the $(q\bar{q})^{(n-\delta)^2/4k}$ factor in the partition function acts like a Gaussian damping those contributions at large $k$.}

The primary dimension \eqref{dimension} takes the finite limit
\begin{align}
\label{scaleddimension}
h_{\{\lambda, n;\delta\}}\, =\, \frac{(\delta -2n)}{4k}\delta\, +\, \frac{\lambda(\lambda+2)}{4(k+2)}\, +\, s_{\lambda,\mu,\nu}\ .
\end{align}
This is finite because in \eqref{dimension} there is a cancellation between divergent terms in $h_{\mu}^{(\ell)}$ and $h_{\nu}^{(k+\ell)}$. Note that the integer shift stays finite in the limit since it depends on the integers $\a$ and $\b$
defined  via  the difference $\n\!-\!\m$ in \eqn{mod2k}.

The limit of the coset character \eqref{cosetcharacter} is simply
\begin{align}
\label{scaledcharacter}
\Xi_{\{\lambda , \mu ; \nu\}}(q)\, &=\, c^{\lambda}_{\beta}(q)\, q^{(n-\delta)^2/4k}\ .
\end{align}
Compared with \eqref{cosetcharacter}, the second term has disappeared - the dimensions of the singular vectors become infinite. Remembering that $n\equiv\nu-\mu$, we have $n=2k\alpha +\beta$ with $-k+1\leqslant \beta \leqslant k$ , hence the square modulus of the character is (we now drop arguments of $\tau$ or $q$ for the sake of brevity)
\be
|\Xi_{\{\lambda , n, \delta\}}|^2\, =\, |c^{\lambda}_{\beta}|^2\, (q\bar{q})^{(\alpha-\frac{\delta}{2k}+\frac{\beta}{2k})^2 k }\ .
\ee
To compute the limit of the partition function \eqref{eq:coset partition function} we need to do the sum over all allowed weights. Remembering the discussion below \eqref{cosetcharacter}, we can do an unconstrained sum over all weights, dividing by 2 to account for \eqref{identification}
\begin{align}
\mathcal{Z}_\mathrm{coset}|_{\ell\rightarrow\infty}\, =\, \frac{1}{2}\sum_{\lambda, \alpha, \beta, \delta}\,|c^{\lambda}_{\beta}|^2\, (q\bar{q})^{(\alpha-\frac{\delta}{2k}+\frac{\beta}{2k})^2 k }\ .
\end{align}
Let's first consider the sum over $\alpha, \delta$ only. The important point to realize is the following: $\alpha$ is an integer and the sum over $0\leqslant\delta\leqslant k$ has infinitesimal spacing $\Delta\delta=k/\ell$, becoming continuous in the limit. We can define $x\equiv \alpha-\frac{\delta}{2k}$, which has spacing $\Delta x=1/(2\ell)$. Looking at the ranges of the parameters, we see that in the limit $x$ takes values in the following intervals, covering half the real line
\begin{align}
x\in \cdots [-\frac{3}{2},-1]\, ,\, [-\frac{1}{2},0]\, ,\, [\frac{1}{2},1]\, ,\, [\frac{3}{2},2]\cdots\ .
\end{align}
In the sum we pull out a factor of $\Delta x$ in order to pass to a Riemann integral
\be
\label{sumtoint}
\begin{split}
&\sum_{\alpha, \delta}|\Xi_{\{\lambda, \alpha, \beta, \delta\}}|^2\ =\  \sum_{\alpha, \delta}\, |c^{\lambda}_{\beta}|^2\, (q\bar{q})^{(\alpha-\frac{\delta}{2k}+\frac{\beta}{2k})^2 k }= 2\ell\,|c^{\lambda}_{\beta}|^2\sum_x\Delta x\, (q\bar{q})^{(x+\frac{\beta}{2k})^2 k }
\\
&\qq\qq\qq \Longrightarrow\ 2\ell\,|c^{\lambda}_{\beta}|^2 \sum_{\alpha\in\mathbb{Z}}\int_{\alpha-\frac{1}{2}}^{\alpha}\mathrm{d}x\, (q\bar{q})^{(x+\frac{\beta}{2k})^2 k }\ .
\end{split}
\ee
After we sum over $-k+1\leqslant\beta\leqslant k$ we can extend the $x$ integral over the whole real axis. To see this, first separate out the $\beta=0, k$ terms. For $\beta=0$ because the integrand is even we can write
\begin{align}
\sum_{\alpha\in\mathbb{Z}}\int_{\alpha-\frac{1}{2}}^{\alpha}\mathrm{d}x\, (q\bar{q})^{x^2 k }\, =\, \sum_{\alpha\in\mathbb{Z}}\int_{\alpha}^{\alpha+\frac{1}{2}}\mathrm{d}x\, (q\bar{q})^{x^2 k }\, =\, \frac{1}{2}\int_{-\infty}^{\infty}\mathrm{d}x\, (q\bar{q})^{x^2 k }
\end{align}
and there is a similar equation for $\beta = k$. The other terms, $-k+1\leqslant\beta\leqslant  -1 $ and $1\leqslant\beta \leqslant k-1$, come in pairs of opposite sign. We can argue the following
\begin{align}
|c^{\lambda}_{\beta}|^2\,&\frac{\ell}{k} \sum_{\alpha\in\mathbb{Z}}\int_{\alpha-\frac{1}{2}}^{\alpha}\mathrm{d}x\, (q\bar{q})^{(x+\frac{\beta}{2k})^2 k }
\, +\, |c^{\lambda}_{-\beta}|^2\,\frac{\ell}{k} \sum_{\alpha\in\mathbb{Z}}\int_{\alpha-\frac{1}{2}}^{\alpha}\mathrm{d}x\, (q\bar{q})^{(x-\frac{\beta}{2k})^2 k }
\nonumber\\
&=\,|c^{\lambda}_{\beta}|^2\,\frac{\ell}{k} \left[\sum_{\alpha\in\mathbb{Z}}\int_{\alpha-\frac{1}{2}}^{\alpha}\mathrm{d}x\, (q\bar{q})^{(x+\frac{\beta}{2k})^2 k }
+\,\sum_{\alpha\in\mathbb{Z}}\int_{\alpha}^{\alpha+\frac{1}{2}}\mathrm{d}x\, (q\bar{q})^{(-x-\frac{\beta}{2k})^2 k }\right]\label{step2}\\
&=\,|c^{\lambda}_{\beta}|^2\,\frac{\ell}{k} \left[\sum_{\alpha\in\mathbb{Z}}\int_{\alpha-\frac{1}{2}}^{\alpha}\mathrm{d}x\, (q\bar{q})^{(x+\frac{\beta}{2k})^2 k }
+\, \sum_{\alpha\in\mathbb{Z}}\int_{\alpha}^{\alpha+\frac{1}{2}}\mathrm{d}x\, (q\bar{q})^{(x+\frac{\beta}{2k})^2 k }\right]\label{step3}\\
&=\,|c^{\lambda}_{\beta}|^2\frac{\ell}{k}\int_{-\infty}^{\infty}\mathrm{d}x\, (q\bar{q})^{(x+\frac{\beta}{2k})^2 k }\, =\, |c^{\lambda}_{\beta}|^2\frac{\ell}{k}\int_{-\infty}^{\infty}\mathrm{d}x\, (q\bar{q})^{x^2 k }\ .
\label{step4}
\end{align}
To get \eqref{step2} we used that $c^{\lambda}_{\beta}=c^{\lambda}_{-\beta}$ and also changed integration variable $x\rightarrow -x$.  For \eqref{step3} we used the symmetry of the integrand in the second term. Finally in \eqref{step4} we have joined the integration domains and shifted $x\rightarrow x-\frac{\beta}{2k}$. Thus the full integral has no $\beta$ dependence at all - it amounted only to a shift in integration variable.

Thus the full sum over $-k+1\leqslant\beta\leqslant k$ gives
\be
\begin{split}
&\sum_{\alpha,\beta, \delta}|\Xi_{\{\lambda, \alpha, \beta, \delta\}}|^2|_{k\rightarrow\infty}= \frac{1}{2}\times\sum_{\beta=-k+1}^k 2\ell\,|c^{\lambda}_{\beta}|^2\, \int_{-\infty}^{\infty}\mathrm{d}x\, (q\bar{q})^{k\, x^2 }
\\ &\qq\qq\qq\qq\quad = \ell\, \sqrt{k}\, \sum_{\beta=-k+1}^k |c^{\lambda}_{\beta}|^2\, \frac{1}{\sqrt{\mathrm{Im}\tau}}\ .
\end{split}
\ee
The final step is to sum over $\lambda$ and divide by 2 to account for the identification \eqref{identification}. The final result for the partition function is
\begin{align}\nonumber
\mathcal{Z}_\mathrm{coset}|_{\ell\rightarrow\infty}\, &=\, \frac{1}{2} \sum_{\alpha,\beta, \delta,\lambda}|\Xi_{\{\lambda, \alpha, \beta, \delta\}}|^2|_{\ell\rightarrow\infty}\\
&=\, \ell\sqrt{k}\,\frac{1}{|\eta (q)|^2}\frac{1}{\sqrt{\mathrm{Im}\tau}}\, \frac{1}{2}\sum_{\lambda =0}^k\sum_{m = -k+1}^k |\eta (q) c^{\lambda}_m|^2\label{dedekind}\\
&= \ell\,\sqrt{k}\,\mathcal{Z}_{\mathrm{free\, boson\, on}\; \mathbb{R}}\, \cdot \, \mathcal{Z}_{k-\mathrm{parafermions}}\label{final}\ .
\end{align}
To arrive at \eqref{dedekind} we pulled out a factor of $|\eta (q)|^{-2}$ from the sum. Up to normalization, \eqref{final} is just the product of the diagonal partition functions of a free uncompactified scalar and the $k$-parafermion coset, corresponding to a known string background and CFT which we can write as $\hat{\mathfrak{su}}(2)_k/\hat{\mathfrak{u}}(1)\times \mathbb{R}$. The result can be understood geometrically as follows. The Hopf fibre of $S^3$ in the WZW model, which had radius $\sim \sqrt{k}$, has decompactified - its radius has been scaled to infinity by a factor of $\ell$. Thus $\ell\,\sqrt{k}$ is the effective radius of the boson after the limit. In contrast, the parafermions have stayed at finite volume.

The background corresponding to this partition function is \eqref{hgeq}. The global twisting does not manifest itself in \eqref{final}, but we can see it by expanding the exponent in \eqref{scaledcharacter}: $\delta$ is the momentum on $\mathbb{R}$, and $n$ is the parafermion quantum number, and there is a cross-term $-2n\delta$ coupling the two. This term gets washed out when we sum over all states in the partition function - it does not affect the spectrum, but changes the OPE. We will see this in section \ref{sec:cosetfusion}.

Finally, note that \eqref{final} is modular invariant, and that it scales linearly with $\ell$, reflecting the fact that the number of weights we have retained is order $\ell$.

\subsubsection{Finite weights}
\label{sec:finite}
In this and the next section we try two other scaling ansatze which can be checked for consistency. Our conclusion will be that \eqref{linearscaling} is the only scaling consistent with modular invariance.

First let us try keeping all weights finite
\begin{align}
\label{finite limit}
\ell\rightarrow\infty \, ,\qquad \l, \m, \n\quad \text{fixed}\ ,
\end{align}
 which means only keeping primary states with dimensions parametrically smaller than $\ell$. The dimension of these primaries is given by the limit \eqref{finite limit} of \eqref{dimension}
\begin{align}
\label{finitedimension}
h_{\{\lambda , \mu ; \nu\}}\, =\, h_{\lambda}^{(k)}\, +\, s_{\lambda,\mu,\nu}\ .
\end{align}
Here $h^{(\ell)}_{\mu}$ and $h^{(k+\ell)}_{\nu}$ have cancelled each other out, so the only difference from the $\hat{\mathfrak{su}}(2)_k$ weights comes from the integer shift.

This limit of the coset character \eqref{cosetcharacter} is
\begin{align}
\label{finitecharacter}
\Xi_{\{\lambda , \mu ; \nu\}}\, =\, c^{\lambda}_{\mu-\nu}\, q^{(\mu-\nu)^2/4k}\, -\, c^{\lambda}_{\mu+\nu-2}\, q^{(\mu+\nu-2)^2/4k}\ .
\end{align}
Note that here, in contrast with \eqref{scaledcharacter}, the singular vectors are retained in the limit.
Substituting \eqref{finitecharacter} into \eqref{eq:coset partition function} gives the limit of the partition function. The form of the limit will depend on exactly how we cut off the sum over $\mu$ and $\nu$, but to take a representative example, let's sum over $0\leqslant \mu, \nu\leqslant w_{\mathrm{max}}$ where $w_{\mathrm{max}}$ is some finite number:
\begin{multline}
\label{finitebothterms}
\mathcal{Z}_\mathrm{coset}|_\mathrm{\ell\rightarrow\infty}\,=\, \sum_{\lambda=0}^k\left(\sum_{n=0}^{2w_{\mathrm{max}}} (w_{\mathrm{max}}-|w_{\mathrm{max}} -n|)|c^{\lambda}_n|^2\, (q\bar{q})^{n^2/4k}\right.
\\ \left. -\, \sum_{\mu,\nu=0}^{w_{\mathrm{max}}} c^{\lambda}_{\mu-\nu}(c^{\lambda}_{\mu+\nu})^* q^{(\mu-\nu)^2/4k}\bar{q}^{(\mu+\nu)^2/4k}\right)\ .
\end{multline}
As $w_{\mathrm{max}}$ becomes large (but still $w_{\mathrm{max}}\ll \ell$) the first term becomes dominant because it scales linearly with $w_{\mathrm{max}}$. This is a discrete sum involving the parafermion characters which can be compared with \eqref{eq:wzw diagonal part function}. All the winding modes have been removed from the free boson, so that
 it is not modular invariant. Thus we conclude that the subset of parametrically small spins is not closed under the action of the modular group. Note the linear scaling of the partition function with $w_{\mathrm{max}}$, as expected from the truncation \eqref{finite limit}.
\subsubsection{Intermediate scaling}
\label{limitthree}
Lastly, let's take a limit similar to \eqref{linearscaling}, but use $\sqrt{\ell}$ instead of $\ell$
\begin{align}
\mu = \frac{\gamma}{k}\, \sqrt{\ell}\qquad \nu = \mu + n\qquad \ell\rightarrow\infty\ ,
\end{align}
where $\gamma$ is treated as finite. Although the primary weights \eqref{dimension} are the same as in the finite case \eqref{finitedimension}, the limiting character resembles \eqref{scaledcharacter} in that there is no subtraction of singular states. It is also independent of $\gamma$
\begin{align}
\Xi_{\{\lambda,\, n,\, \gamma\}}\, &=\, c^{\lambda}_n\, q^{n^2/4k} \ .
\end{align}

The partition function \eqref{eq:coset partition function} becomes
\begin{align}
\label{sqrtpartition}
\mathcal{Z}_\mathrm{coset}|_{\ell\rightarrow\infty}\, =\, \frac{1}{2}\gamma_{\mathrm{max}}\sqrt{\ell}\sum_{\lambda=0}^k\, \sum_{n\in \mathbb{Z}}|c_n^{\lambda}|^2\, (q\bar{q})^{n^2/4k}\ ,
\end{align}
where $\gamma_{\mathrm{max}}$ is an order one cutoff on the sum over $\gamma$, analogous to $w_{\rm max}$ in section \ref{sec:finite}. This is equal to the WZW partition function but with winding modes removed, and is thus not modular invariant.

Note that, by breaking up the sum over $n$ modulo $2k$ as in \eqref{mod2k} we can see that the spectrum of chiral dimensions in this limit is identical to the WZW theory. Indeed we can rearrange the limiting coset characters into $\hat{\mathfrak{su}}(2)_k$ characters
\begin{align}
\sum_{n\in\mathbb{Z}} \Xi_{\{\lambda,\, n,\, \gamma\}}= \sum_{\alpha,\beta\in\mathbb{Z}} \Xi_{\{\lambda,\, \alpha,\, \beta,\, \gamma\}} = \sum_{\beta = -k+1}^k \left( c^{\lambda}_{\beta} \sum_{\alpha\in\mathbb{Z}} q^{(\alpha+\frac{\beta}{2k})^2 k }\right) = \chi^{(k)}_{\lambda} (q)\  .
\end{align}
Thus we can recover the original WZW partition function as a certain non-diagonal modular invariant in this limit. However, note that this would not correspond to a direct quantization of the gauged WZW action. We are not sure if this observation has any deeper significance.
\subsection{Modular matrix}
The modular transformations for the coset are given by \eqref{eq:T coset}. We are interested in the limiting behavior,
as $\ell\to \infty$, of the last two factors of $\mathcal{S}$, containing the $\ell$ dependence.
As before, we want to see at the level of power counting in $\ell$ how the truncation to weights \eqref{linearscaling} is consistent with modular invariance. We again use the schematic notation of section \ref{sec: wzwmod} to represent the scaling of the various quantities. That is, the number of weights with linear scaling is $n_{\ell}$ and the coset characters are of order $\Xi_{\ell}$.

The key feature of \eqref{linearscaling} is that the weights $\mu$ and $\nu$ are allowed to scale linearly in $\ell$, but they must have finite difference. That is, the leading linear term in each is the same:
\begin{align}
\label{finitediff}
\mu &\sim \delta\ell+\cdots \ ,\qq \nu \sim \delta\ell+\cdots\  .
\end{align}
The subtlety in the power counting is that $\mu-\nu$ is kept fixed while $\mu$ can range up to $\ell$. Therefore the number of weights $\mu,\nu$ within the limit ansatz \eqref{linearscaling} is $n_{\ell}=\mathcal{O}(\ell)$ and not $\mathcal{O}(\ell^2)$ as one would na\"{i}vely think from a double sum over $\mu$ and $\nu$. The characters are order one, i.e. $\Xi_{\ell}=\mathcal{O}(1)$.

First look at the computation of the partition function \eqref{eq:coset partition function}
\begin{align}
\mathcal{Z}_{\rm coset}\, =\, \sum_{\mu,\nu}|\Xi_{\mu,\nu}|^2\, \sim\, n_{\ell}\Xi_{\ell}^2\, =\, \ell\cdot 1^2\, =\, \ell\ ,
\end{align}
so the linear weights give the correct scaling, as in \eqref{final}.

Now consider the transformation of the partition function while only summing over states in the scaling ansatz. The modular matrix between two states will be
\begin{align}
\mathcal{S}_{\delta\delta'}^{\rm coset}\, \sim\, \mathcal{S}_{\delta\delta'}^2\sim \frac{1}{\ell}\sin^2\left( \ell(\delta\delta') \right)\ ,
\end{align}
because of \eqref{finitediff}. Therefore the transformed partition function scales as
\be
\label{samescaling}
\begin{split}
& \mathcal{Z}_{\rm coset}\left( -\frac{1}{\tau} \right) = \sum_{\mu,\nu}|\sum_{\mu',\nu'}\mathcal{S}_{\mu,\mu'}\mathcal{S}_{\nu,\nu'}\Xi_{\mu',\nu'}|^2
\\
&\qq\qq\qq \sim\, n_{\ell} \left( n_{\ell}\Xi_{\ell}\, \mathcal{S}_{\delta\delta'}^{\rm coset} \right)^2
\\
&\qq\qq\qq \sim\ell \sin^4(\delta\delta'\ell)\sim \ell\ .
\end{split}
\ee
In the last stage we used that the integral of the factor involving the $\sin$-function will be some positive order one number. The correct scaling is recovered by staying within the scaling ansatz, so the truncation is consistent with modular invariance.
\subsection{Fusion rules}
\label{sec:cosetfusion}
As in the WZW model, the Verlinde formula allows us to check whether the high spin limit \eqref{linearscaling} can be consistent. We use the coset $\mathcal{S}$ matrix \eqref{eq:T coset} in the Verlinde formula \eqref{verlinde}, where now we must do a triple sum over all coset fields. Again, since $k$ is finite the sum over $\lambda$ does not play a role in our argument and we leave it out.

We want to check the following: given two coset fields $(\mu_1,\nu_1)\sim(\delta_1\ell,\delta_1\ell)$ and $(\mu_2,\nu_2)\sim(\delta_2\ell,\delta_2\ell)$ with scaling \eqref{finitediff}, and in \eqref{verlinde} summing only over weights in the ansatz, are the fusion coefficients always zero for a generic third field $(\mu_3,\nu_3)\sim(\tilde{\mu}_3\ell,\tilde{\nu}_3\ell)$ outside the ansatz, i.e. where $\mu_3$ and $\nu_3$ scale independently with $\ell$? To implement the truncation, the key thing we do is to write the double sum over $\mu,\nu$ as a single integral over $x$. Then we find
\begin{align}
\mathcal{N}_{(\mu_1,\nu_1), (\mu_2,\nu_2)}^{\phantom{(\mu_1,\nu_1)\;\;\;} (\mu_3,\nu_3)}|_{\ell\rightarrow\infty}^{\mathrm{trunc.}}& \propto
\frac{1}{\ell}\int_0^1\dd x\, \frac{\sin^2 (\pi\delta_1\ell\, x)\, \sin^2 (\pi\delta_2\ell\, x)}{\sin^2 (\pi x)}\sin (\pi\tilde{\mu}_3\ell\, x) \sin (\pi\tilde{\nu}_3\ell\, x)
\nonumber\\
=\quad \frac{1}{16}&\sum s_2 s_3|\delta_1 + s_1 \delta_2+\frac{1}{2}(s_2\tilde{\mu}_3+s_3\tilde{\nu}_3)|\nonumber\\
-\frac{1}{8}&\sum s_1 s_2|\delta_1+\frac{1}{2}(s_1\tilde{\mu}_3+s_2\tilde{\nu}_3)|-\frac{1}{8}\sum s_1 s_2|\delta_2+\frac{1}{2}(s_1\tilde{\mu}_3+s_2\tilde{\nu}_3)|\nonumber\\
+\frac{1}{4}&\sum s_1|\frac{1}{2}(\tilde{\mu}_3+s_1\tilde{\nu}_3)|\quad .
\label{eq:cosetfusion}
\end{align}
where the sums are over all combinations of signs $s_i=\pm 1$. Note that, whereas in the coset before the limit the fusion coefficients were finite natural numbers, once we truncate the sum in the Verlinde formula \eqref{verlinde} the fusion rules become a continuous function - as a clarifying point, note that the theory is no longer rational with respect to the coset algebra. Interestingly, \eqref{eq:cosetfusion} does \textit{not} vanish for $\tilde{\mu}_3\neq\tilde{\nu}_3$, suggesting that the weights we have kept are not closed under the operator product expansion. This is in contrast to the situation in the partition function. There, weights outside the truncation can be
consistently ignored because their conformal dimension becomes large with $\ell$, and it can be seen from eqs. \eqref{eq:coset partition function} \eqref{cosetcharacter}, that their contribution is exponentially suppressed. In the fusion rules however they seem to play an important role.

Thanks to the geometric result \eqref{final}, the quantum numbers $\delta_i$ above have the interpretation as momenta along the decompactified Hopf fibre. We can ask the following question: are the truncated fusion rules the same as the product CFT $\hat{\mathfrak{su}}(2)_k/\hat{\mathfrak{u}}(1)\times \mathbb{R}$? The answer is no. If $\delta_i$ were the momenta along a decoupled $\mathbb{R}$ factor in the geometry we would have $\mathcal{N}_{\delta_1\delta_2}^{\phantom{\delta_1\delta_2}\delta_3}=0$ unless $\sum\delta_i=0$, i.e. momentum would be conserved, but instead we get the rather non-trivial result \eqref{eq:cosetfusion}. This is possible precisely because of the cross-term mentioned in the discussion after \eqref{final}. In fact, we should think of $n$ and $\delta$ as momenta along the maximal tori of two $SU(2)$'s, and there can be momentum transfer because of the coupling between them. Therefore parafermions do not factor out in the fusion rules. The fact that the OPE differs from $\mathbb{R}$ was already noted in the $k=1$ case \cite{Runkel:2002yb}.

\section{Comparison with the Abelian case}
\label{sec:abelian}

According to the general proposal, the Abelian T-dual of the $\hat{\mathfrak{su}}(2)_k$ WZW model is given by the $\ell\rightarrow\infty$ of the coset $\hat{\mathfrak{su}}(2)_k\oplus \hat{\mathfrak{u}}(1)_{\ell} / \hat{\mathfrak{u}}(1)_{k+\ell}$.

The calculations closely parallel the non-Abelian case, so we will not give details here. The relevant character decomposition is
\begin{align}
\chi^{(k)}_{\lambda}(q,z)\, \zeta^{(\ell)}_{\mu}(q,z)\, =\, \sum_{\nu}\Xi_{\{\lambda,\mu;\nu\}}^\mt{Abelian}(q)\, \zeta^{(k+\ell)}_{\nu}(q,z)\ ,
\end{align}
where we have introduced the $\hat{\mathfrak{u}}(1)_k$ characters $\zeta^{(k)}_{\mu}(q,z)\equiv \Theta^{(k)}_{\mu} (q,z)/\eta (q)$. $\lambda$ is the highest weight for a $\hat{\mathfrak{su}}(2)_k$, whereas $\mu$ and $\nu$ are now the labels of $\hat{\mathfrak{u}}(1)_{\ell}$ and $\hat{\mathfrak{u}}(1)_{k+\ell}$ representations respectively, so that $0\leqslant\lambda\leqslant k$ and $-\ell +1\leqslant\mu\leqslant\ell$, $-(k+\ell) +1\leqslant\nu\leqslant k+\ell$.  Then when $\ell\rightarrow\infty$, $\mu$ and $
\nu$ can take all integer values. Using this we can compute $\Xi_{\{\lambda,\mu;\nu\}}^\mt{Abelian}(q)$ by a derivation similar to the one in appendix \ref{characterderivation} for the non-Abelian coset. The result is
\begin{align}
\Xi_{\{\lambda,\mu;\nu\}}^\mt{Abelian}(q) = c_{\m-\nu}^{\lambda}\, q^{ \frac{(\mu-\nu)^2}{4k} + \frac{\mu^2}{4\ell} - \frac{\nu^2}{4(k+\ell)}}\ ,
\end{align}
which is similar to \eqref{cosetcharacter}, but with the second term removed and without any shift in the levels. The partition function is now
\begin{align}
\label{abelianz}
\mathcal{Z}_\mt{Abelian}(q) = \sum_{\lambda,\mu,\nu}|\Xi_{\{\lambda,\mu;\nu\}}^\mt{Abelian}(q)|^2\quad \ ,
\end{align}
subject to the usual selection rule $\lambda+\mu-\nu\in 2\mathbb{Z}$ and field identification $\{\lambda,\mu,\nu\}\sim \{k-\lambda,\ell-\mu,(k+\ell)-\nu\}$.
Now we take the limit analogous to \eqref{linearscaling}
\begin{align}
\label{abelianscaling}
\mu\equiv\frac{\delta}{k}\, \ell\ , \quad \nu-\mu\equiv n\ , \quad -k<\delta\leqslant k
\end{align}
and obtain up to a constant factor the same partition function as in the non-Abelian case
\begin{align}
\label{abelianfinal}
\mathcal{Z}_\mt{Abelian}|_{\ell\rightarrow\infty}\, = \, 2\, \ell\sqrt{k}\,\mathcal{Z}_{\mathrm{free\, boson\, on}\; \mathbb{R}} \, \mathcal{Z}_{k-\mathrm{parafermions}}\ ,
\end{align}
where the extra factor of 2 compared with \eqref{final} comes from the different range for $\delta$ in \eqref{abelianscaling} compared with \eqref{linearscaling}. This is modular invariant, and is not a surprising result given that the Abelian dual geometry to the WZW background \eqref{backwzw} is in fact locally the same as \eqref{hgeq}. However, \eqref{abelianfinal} is not the partition function of the Abelian T-dual - the Abelian duality is exact, so the dual partition function should be equal to $\mathcal{Z}_\mt{WZW}$.

The reason for the discrepancy is that in the `real' Abelian dual background the coordinate $z$ in \eqref{hgeq} is compactified with period $2\pi$. This means we must perform an orbifolding procedure on \eqref{abelianz}: subtract all the non-periodic modes, and then add twisted sectors. After that we find that the partition is indeed modified to be equal to the original $\hat{\mathfrak{su}}(2)_k$ model. This is the standard argument appearing for example in \cite{Rocek:1991ps}, applied to the WZW model. However, from the perspective of the coset these twisted modes never appear - they must be added by hand. This is a deficiency of the coset approach in the Abelian setting, but in the analogous non-Abelian case there is no obvious way to compactify, so the coset method is as good as any other.
\section{Discussion}
\label{sec:discussion}

We presented the partition function corresponding to the non-Abelian T-dual  of the $\hat{\mathfrak{su}}(2)_k$ WZW model and found that it is equal to the product of the partition functions of parafermions and a free boson. Because in this case the proposal \eqref{eq:coset su2 diagonal} relates the duality to a procedure defined purely in CFT, it is guaranteed that conformal invariance is preserved. The coset construction allowed us to make exact analytic statements. It should be possible to
extend our analysis to other groups, where similar CFT techniques are applicable, though this is expected to be technically more
involved.

The dual partition function \eqref{final} is obviously not the same as the original WZW partition function. This is related to the non-compactness of the dual target space, and so is not a surprise. Indeed the modern view on non-Abelian duality is as a map between different string backgrounds.

In the large level limit of the coset two different backgrounds \eqref{hgeq} and \eqref{non1} appear to share the same string spectrum and partition functions.
While we were able to argue this indirectly, it is desirable to make the proposal more concrete with direct reference to the target space
as was done in \cite{Polychronakos:2010fg} for the dilaton mode.

It can be shown that the backgrounds \eqref{hgeq} and \eqref{non1} are not diffeomorphic equivalent to each other. Given this, a natural question to ask is whether strings quantized on the backgrounds are actually equivalent to each other. There are two reasons one might doubt this conclusion.
Firstly, in order to prove equivalence of the theories we would need to show equality not only of the partition function, i.e. the spectrum, but also of the operator product expansions of the states in the theories.  Secondly, what we have done in this paper is to first quantize strings on the coset background \eqref{cosetmetric}, \eqref{cosetfields} and then take the limit in the quantum theory.
This is not necessarily equivalent to quantizing  \eqref{hgeq} and \eqref{non1} directly. There is no a priori reason why these two procedures -- quantization and the large-$\ell$ limit -- should commute. A test of the equivalence would involve working directly with the two limiting backgrounds, which may be technically difficult because of the lack of group theoretic structure. This is beyond the scope of this paper, but would be an interesting direction to follow up.

Finally, it will be interesting to see if our findings may have  relevance in the context of non-Abelian T-duality in the
presence of Ramond-Ramond backgrounds \cite{Sfetsos:2010uq}.

\subsection*{Acknowledgments}
We would like to thank Daniel Thompson, who suggested to us the relevance of parafermions. BF appreciates an enjoyable stay at ETH Zurich, where he had useful discussions with Ben Hoare. We also thank the theory group at CERN, where some of the work was completed. This research is partly supported by the "CUniverse" research promotion project by Chulalongkorn University (grant reference CUAASC).

\appendix

\section{Factorized form of \texorpdfstring{$\hat{\mathfrak{su}}(2)_k$}{su(2)} WZW theory}
\label{factorized_su2}
First let us define the $\hat{\mathfrak{su}}(2)_k$ character in a highest weight representation $R_{\lambda}$
\begin{align}
\chi^{(k)}_{\lambda}(q,z)\, \equiv\, \mathrm{Tr}_{R_{\lambda}} q^{L_0}z^{2 J_0^3}
\end{align}
and the $\hat{\mathfrak{su}}(2)_k$ generalized theta functions
\begin{align}
\label{thetafns}
\Theta^{(k)}_m (q,z)\, \equiv \, \sum_{n\in\mathbb{Z}}\, q^{k(n+\frac{m}{2k})^2} z^{m+2k\, n}\quad .
\end{align}
We obtain their `specialized' versions by setting $z=1$:
\begin{align}
\chi^{(k)}_{\lambda}(q)\, &\equiv\, \chi^{(k)}_{\lambda}(q,1)& \Theta^{(k)}_m (q)\, &\equiv \, \Theta^{(k)}_m (q,1)\quad .
\end{align}

In this paper we use the following rewriting of the $\hat{\mathfrak{su}}(2)_k$ characters \cite[Chap.\, 18]{DiFrancesco:1997nk}
\begin{align}
\label{factorizedcharacter}
\chi^{(k)}_{\lambda}(q,z)\, =\, \sum_{m=-k+1}^{k}\, c^{\lambda}_m (q)\, \Theta^{(k)}_m (q,z)
\end{align}
where $c^{\lambda}_m (q)$ are
\begin{align}
\label{stringnormalization}
&c^{\lambda}_m (q)\, \equiv \,
\begin{cases}
q^{h^{\lambda}_m (\mathrm{parafermion})-\frac{c}{24}}\,\sigma^{\lambda}_m (q) & -\lambda\leq m\leqslant \lambda\\
c^{k-\lambda}_{k-m} & \lambda<|m|\leqslant k\\
c^{\lambda}_{m - 2k} & |m|>k
\end{cases}\nonumber\\
&h^{\lambda}_m (\mathrm{parafermion})\, \equiv\, h_{\lambda}^{(k)}-\frac{m^2}{4k}
\end{align}
By design we have the shift and reflection symmetries $c^{\lambda}_{m} = c^{\lambda}_{m + 2k} = c^{\lambda}_{k-m}$. The string functions $\sigma^{\lambda}_m$ have their $q^0$ coefficient normalized to one $\sigma (q)=1+\#q+\cdots$, and generate the multiplicities of states in the strings within highest weight representations (see \cite[Chap.\, 14]{DiFrancesco:1997nk}). They have an explicit expression derived in \cite{Distler:1989xv}:
\begin{align}
\label{paracharacters}
\sigma^{\lambda}_m (q)\, =\, &\frac{q^{\frac{1}{4(k+2)}+\frac{c}{24}}}{\eta (q)^3}\sum_{r,s=0}^{\infty}\, (-1)^{r+s}\, q^{\frac{1}{2}r(r+1)+\frac{1}{2}s(s+1)+rs(k+1)}\hspace{0.5in}\\
&\hspace{1in}\times\left[ q^{s\frac{\lambda-m}{2}+r\frac{\lambda+m}{2}}\, -\, q^{k+1-\lambda+s(k+1-\frac{\lambda-m}{2})+r(k+1-\frac{\lambda+m}{2})} \right]\nonumber\\
&\hspace{2in}\mathrm{for}\;\lambda - m \in 2\mathbb{Z}\nonumber\\
\sigma^{\lambda}_m (q)\, =\; &0\quad \mathrm{otherwise}.\nonumber
\end{align}
Note the selection rule $\sigma^{\lambda}_m (q)=0$ for $\lambda - m \notin 2\mathbb{Z}$, which is useful when writing the coset partition function.

We have verified in Mathematica (up to sufficiently high order in $q$) that inserting \eqref{paracharacters} and \eqref{stringnormalization} into the sum \eqref{factorizedcharacter} reproduces \eqref{su2characteroriginal} in the main text.

We can further write \eqref{factorizedcharacter} as
\begin{align}
\chi^{(k)}_{\lambda}(q,z)\, =\, \sum_{m=-k+1}^{k} \eta (q) c^{\lambda}_m (q)\; \left(\frac{\Theta^{(k)}_m (q,z)}{\eta (q)}\right)
\end{align}
where $\eta (q)\,\equiv\, q^{\frac{1}{24}}\prod_{n=1}^{\infty}(1-q^n)$ is the Dedekind eta function. This shows the decomposition of $\hat{\mathfrak{su}}(2)_k$ characters with respect to $\hat{\mathfrak{u}}(1)_k$. The quantity in brackets is a $\hat{\mathfrak{u}}(1)_k$ character, the current algebra for a free boson compactified on a circle of radius $\sqrt{2k}$\;\footnote{We use the normalization $S=1/8\pi \int \partial_{\mu}\phi\partial^{\mu}\phi$, $\phi\sim \phi +2\pi R$}, and the prefactor is a character for the coset $\hat{\mathfrak{su}}(2)_k / \hat{\mathfrak{u}}(1)$, sometimes known as the parafermion theory. This is the reason for the notation in \eqref{stringnormalization} -  $h^{\lambda}_m (\mathrm{parafermion})$ are the conformal weights of the parafermion primary fields.

This means we can think of the WZW model as being $k$-parafermions coupled to a compactified boson. Intuitively, this circle corresponds to the maximal torus of the $SU(2)$ group manifold. When $k=1$ we have $\eta (q) c^{\lambda}_m=\delta_m^{\lambda}$, so the $\hat{\mathfrak{su}}(2)_1$ theory is a free boson at the self-dual radius $R=\sqrt{2}$.

\section{Large \texorpdfstring{$\ell$}{l} limit of coset characters}
\label{characterderivation}
Rather than deriving an expression for the coset character at finite $\ell$ (as in e.g.\cite[Chap. 18]{DiFrancesco:1997nk}) and then taking $\ell\rightarrow\infty$, it is possible to derive them directly at large $\ell$.

The coset characters are defined by the branching formula \eqref{eq:char decomposition id}, which we repeat here for convenience
\begin{align}
\label{char_dec_appendix}
\chi^{(k)}_{\l}(q,z)\chi^{(\ell)}_{\mu}(q,z) = \sum_{\nu=0}^{k+\ell}\Xi_{\{\l,\mu;\nu\}}(q)\chi^{(k+\ell)}_{\nu}(q,z)\quad .
\end{align}

The Kac-Weyl formula gives another expression for the characters in terms of the generalized theta functions:
\begin{align}
\label{kacweyl}
\chi^{(k)}_{\lambda}(q,z)\, &=\, \frac{\Theta^{(k+2)}_{\lambda+1} (q,z)\, -\, \Theta^{(k+2)}_{-\lambda-1} (q,z)}{\Theta^{(2)}_1 (q,z)\, -\, \Theta^{(2)}_{-1} (q,z)}
\end{align}
At large level the theta functions simplify - only the $n=0$ term in \eqref{thetafns} survives:
\begin{align}
\label{largektheta}
\Theta^{(\ell)}_m (q,z)|_{\ell\rightarrow\infty}\, =\, q^{m^2/4\ell} z^m
\end{align}
and so \eqref{kacweyl} becomes
\begin{align}
\label{kacweyllim}
\chi^{(\ell)}_{\mu}(q,z)|_{\ell\rightarrow\infty}\, &=\, q^{\frac{(\mu+1)^2}{4\ell}}\frac{z^{\lambda+1}-z^{-\lambda-1}}{\Theta^{(2)}_1 (q,z)\, -\, \Theta^{(2)}_{-1} (q,z)}\nonumber\\
&=\, q^{\frac{(\mu+1)^2}{4\ell}}\frac{\mathrm{ch}_{\mu+1}(z)}{(z-\frac1z)^{-1}(\Theta^{(2)}_1 (q,z)\, -\, \Theta^{(2)}_{-1} (q,z))}
\end{align}
where $\mathrm{ch}_{\mu}\equiv\frac{z^{\mu+1}-z^{-\mu-1}}{z-z^{-1}}$ is the character of the finite algebra $\mathfrak{su}(2)$.

In taking the limit of \eqref{char_dec_appendix}, $\chi^{(\ell)}_{\mu}(q,z)$ and $\chi^{(k+\ell)}_{\nu}(q,z)$ have large level, and we use formula \eqref{kacweyllim}. Then, after cancelling common denominators from each side of the equation, \eqref{char_dec_appendix} reads
\begin{align}
q^{\frac{(\mu+1)^2}{4\ell}}\chi^{(k)}_{\lambda}(q,z)\, \mathrm{ch}_{\mu+1}(z)\, =\, \sum_{\nu=0}^l\, q^{\frac{(\nu+1)^2}{4(k+\ell)}}\Xi_{\{ \lambda, \mu ; \nu \}}(q)\, \mathrm{ch}_{\nu+1}(z)\quad .
\end{align}
Now we use the orthogonality of the weights with respect to the scalar product:
\begin{align}\label{eq:orthogonality}
\left(-\frac{1}{2}\right)\frac{1}{2\pi\mathrm{i}}\ointctrclockwise\frac{\mathrm{d}z}{z}\left(z-\frac1z\right)^2\, \mathrm{ch}_{\lambda}(z)\mathrm{ch}_{\mu}(z)\, =\, \delta_{\lambda,\mu}
\end{align}
where the contour is around the origin $z=0$. We find
\begin{align}
\Xi_{\{ \lambda, \mu ; \nu \}}\, =\, q^{\frac{(\mu+1)^2}{4\ell}}q^{-\frac{(\nu+1)^2}{4(k+\ell)}}\left(-\frac{1}{2}\right)\frac{1}{2\pi\mathrm{i}}\ointctrclockwise\frac{\mathrm{d}z}{z}\,\left(z-\frac1z\right)^2\, \chi^{(k)}_{\lambda}(q,z)\, \mathrm{ch}_{\mu+1}(z)\, \mathrm{ch}_{\nu+1}(z).
\end{align}

After substituting the expression \eqref{factorizedcharacter} for the (finite $k$) character in terms of parafermions, the integrand has a simple rational form, and the contour integral just picks out the simple poles. To simplify expressions we define $a\equiv\mu +1$, $b\equiv\nu +1$. Also, in substituting \eqref{thetafns},\eqref{factorizedcharacter} we consider the sum over $n$ as an extension of the range of summation of $m$, thus combining the double sum over $n$ and $m$ into a single one:
\begin{align}
\label{contour}
\Xi_{\{\lambda,\mu ;\nu\}}(q)\, &=\, \left(-\frac{1}{2}\right)\frac{1}{2\pi\mathrm{i}}\ointctrclockwise\frac{\mathrm{d}z}{z}\, \sum_{m\in\mathbb{Z}} c_m^{\lambda} q^{ \frac{m^2}{4k} + \frac{a^2}{4\ell} - \frac{b^2}{4(k+\ell)}}\, z^m\, ( z^a-z^{-a})(z^b-z^{-b})\nonumber\\
&=\, \left(-\frac{1}{2}\right)\frac{1}{2\pi\mathrm{i}}\ointctrclockwise\frac{\mathrm{d}z}{z}\, \sum_{m\in\mathbb{Z}} c_m^{\lambda} q^{ \frac{m^2}{4k} + \frac{a^2}{4\ell} - \frac{b^2}{4(k+\ell)}}\left[ z^{a+b+m}+z^{-a-b+m} \, -\, (b\rightarrow -b)\right]\nonumber\\
&=\, c_{a-b}^{\lambda} q^{ \frac{(a-b)^2}{4k} + \frac{a^2}{4\ell} - \frac{b^2}{4(k+\ell)}}\, -\, (b\rightarrow -b)\nonumber\\
&=\, c_{\m-\nu}^{\lambda}\, q^{ \frac{(\mu-\nu)^2}{4k} + \frac{(\mu+1)^2}{4\ell} - \frac{(\nu+1)^2}{4(k+\ell)}}\, -\, c_{\mu+\nu - 2}^{\lambda}\, q^{ \frac{(\mu-(2-\nu))^2}{4k} + \frac{(\mu+1)^2}{4\ell} - \frac{((2-\nu)+1)^2}{4(k+\ell)}}
\end{align}
which is a rewriting of \eqref{cosetcharacter}.

\small

\newcommand\arxiv[2]      {\href{http://arXiv.org/abs/#1}{#2}}
\newcommand\doi[2]        {\href{http://dx.doi.org/#1}{#2}}
\newcommand\httpurl[2]    {\href{http://#1}{#2}}


\begin{thebibliography}{AAA}
\setlength{\itemsep}{-.1em}

\bibitem{Gepner:1986wi}
D.~Gepner and E.~Witten,
{\it String Theory on Group Manifolds},\hfill\break
\doi{10.1016/0550-3213(86)90051-9}{Nucl. Phys. {\bf B278} (1986) 493}.

\bibitem{Witten:1983ar}
E.~Witten,
{\it Nonabelian Bosonization in Two-Dimensions},\hfill\break
\doi{10.1007/BF01215276}{Commun.\ Math.\ Phys.\ {\bf 92} (1984) 455}.

\bibitem{Rocek:1991ps}
M.~Ro\v{c}ek and E.~P.~Verlinde,
{\it Duality, Quotients, and Currents},\hfill\break
\doi{10.1016/0550-3213(92)90269-H}{Nucl.\ Phys. {\bf B373} (1992) 630},
\arxiv{arXiv:hep-th/9110053}{arXiv:9110053 [hep-th]}.

\bibitem{Giveon:1993ai}
  A.~Giveon and M. Rocek,
  {\it On Nonabelian Duality},\hfill\break
  \doi{10.1016/0550-3213(94)90230-5}{Nucl.Phys. {\bf B421} (1994) 173-190},
  \arxiv{arXiv:hep-th/9308154}{arXiv:9308154 [hep-th]}.

\bibitem{Sfetsos:1994vz}
K.~Sfetsos,
{\it Gauged WZW Models and Nonabelian Duality},\hfill\break
\doi{10.1103/PhysRevD.50.2784}{Phys. Rev.  {\bf D50} (1994) 2784},
\arxiv{arXiv:hep-th/9402031}{arXiv:9402031 [hep-th]}.

\bibitem{Alvarez:1994np}
E.~\'Alvarez, L.~\'Alvarez-Gaum\'e and Y.~Lozano,
{\it On Nonabelian Duality},\hfill\break
\doi{10.1016/0550-3213(94)90093-0}{Nucl. Phys. {\bf B424} (1994) 155},
\arxiv{arXiv:hep-th/9403155}{arXiv:9403155 [hep-th]}.

\bibitem{Gaberdiel:1995mx}
M.~R.~Gaberdiel,
{\it Abelian Duality in WZW Models},\hfill\break
\doi{10.1016/0550-3213(96)00181-2}{Nucl.\ Phys. {\bf B471} (1996) 217},
\arxiv{arXiv:hep-th/9601016}{arXiv:9601016 [hep-th]}.

\bibitem{Klimcik:1995ux}
C.~Klimcik and P.~Severa,
{\it Dual Nonabelian Duality and the Drinfeld Double},\hfill\break
\doi{10.1016/0370-2693(95)00451-P}{Phys.\ Lett. {\bf B351} (1995) 455},
\arxiv{arXiv:hep-th/9502122}{arXiv:9502122 [hep-th]}.

\bibitem{Alekseev:1995ym}
A.~Y.~Alekseev, C.~Klimcik and A.~A.~Tseytlin,
{\it Quantum Poisson-Lie T Duality and WZNW Model},
\doi{10.1016/0550-3213(95)00575-7}{Nucl.\ Phys. {\bf B458} (1996) 430},
\arxiv{arXiv:hep-th/9509123}{arXiv:9509123 [hep-th]}.


\bibitem{Klimcik:1996nq}
C.~Klimcik and P.~Severa,
{\it Nonabelian Momentum Winding Exchange},\hfill\break
Phys. Lett. {\bf B383} (1996) 281
\doi{10.1016/0370-2693(96)00755-1},
\arxiv{arXiv:hep-th/9605212}{arXiv:9605212 [hep-th]}.


\bibitem{Klimcik:1996hp}
C.~Klimcik and P.~Severa,
{\it Open Strings and D-Branes in WZNW Model},\hfill\break
\doi{10.1016/S0550-3213(97)00029-1}{Nucl. Phys. {\bf B488} (1997) 653},
\arxiv{arXiv:hep-th/9609112}{arXiv:9609112 [hep-th]}.



\bibitem{Goddard:1984vk}
P.~Goddard, A.~Kent and D.~I.~Olive,
{\it Virasoro Algebras and Coset Space Models},\hfill\break
\doi{10.1016/0370-2693(85)91145-1}{Phys. Lett. {\bf 152B} (1985) 88}.

\bibitem{DiFrancesco:1997nk}
  P.~Di Francesco, P.~Mathieu, D.~S\'{e}n\'{e}chal,
{\it Conformal Field Theory},
Springer, 1996.



\bibitem{Polychronakos:2010fg}
A.~P.~Polychronakos and K.~Sfetsos,
{\it Solving Field Equations in Non-Isometric Coset CFT Backgrounds},
\doi{10.1016/j.nuclphysb.2010.08.001}{Nucl. Phys. {\bf B840} (2010) 534},
\arxiv{arXiv:1006.2386}{arXiv:1006.2386 [hep-th]}.

\bibitem{Polychronakos:2010hd}
A.~P.~Polychronakos and K.~Sfetsos,
{\it High Spin Limits and Non-Abelian T-Duality},
\doi{10.1016/j.nuclphysb.2010.09.006}{Nucl. Phys. {\bf B843} (2011) 344},
\arxiv{arXiv:1008.3909}{arXiv:1008.3909 [hep-th]}.

\bibitem{Alvarez:1994dn}
E.~\'Alvarez, L.~\'Alvarez-Gaum\'e and Y.~Lozano,
{\it An Introduction to T Duality in String Theory},
\doi{doi:10.1016/0920-5632(95)00429-D}{Nucl. Phys. Proc. Suppl. {\bf 41} (1995) 1},
\arxiv{arXiv:hep-th/9410237}{arXiv:9410237 [hep-th]}.


\bibitem{Fredenhagen:2014kia}
  S.~Fredenhagen and C.~Restuccia,
  {\it The Large Level Limit of Kazama-Suzuki Models},
  \arxiv{arXiv:1408.0416}{arXiv:1408.0416 [hep-th]}.

\bibitem{Gaberdiel:2011aa}
  M.~R.~Gaberdiel and P.~Suchanek,
  {\it Limits of Minimal Models and Continuous Orbifolds},
  \doi{10.1007/JHEP03(2012)104}{JHEP {\bf 03} (2012) 104},
  \arxiv{arXiv:1112.1708}{arXiv:1112.1708 [hep-th]}.

\bibitem{Gaberdiel:2014vca}
  M.~R.~Gaberdiel and M.~Kelm,
  {\it The Continuous Orbifold of $\mathcal{N}=2$ Minimal Model\\ Holography},
  \doi{10.1007/JHEP08(2014)084}{JHEP {\bf 08} (2014) 084},
  \arxiv{arXiv:1406.2345}{arXiv:1406.2345 [hep-th]}.

\bibitem{Restuccia:2013tba}
  C.~Restuccia,
  {\it Limit theories and continuous orbifolds},
  \arxiv{arXiv:1310.6857}{arXiv:1310.6857 [hep-th]}.

\bibitem{Runkel:2002yb}
  I.~Runkel and G.M.T.~Watts,
  {\it A non-rational CFT with central charge 1},\hfill\break
  \doi{10.1002/1521-3978(200209)50:8/9<959::AID-PROP959>3.0.CO;2-\%23}{Fortsch.Phys. {\bf 50} (2002) 959-965},
  \arxiv{arXiv:hep-th/0201231}{arXiv:0201231 [hep-th]}.

\bibitem{Roggenkamp:2003qp}
D.~Roggenkamp and K.~Wendland
{\it Limits and degenerations of unitary conformal field theories},
\doi{10.1007/s00220-004-1131-6}{Commun.Math.Phys. {\bf 251} (2004) 589-643},
\arxiv{arXiv:hep-th/0308143}{arXiv:0308143 [hep-th]}.

\bibitem{Verlinde:1988sn}
  E.P.~Verlinde,
 {\it Fusion Rules and Modular Transformations in 2D Conformal Field Theory},
  \doi{10.1016/0550-3213(88)90603-7}{Nucl. Phys. {\bf B300} (1988) 360}.

\bibitem{Bardacki:1990wj}
K.~Bardakci, M.~J.~Crescimanno and E.~Rabinovici
{\it Parafermions from Coset Models},\hfill\break
\doi{10.1016/0550-3213(90)90365-K}{Nucl.\ Phys. {\bf B344} (1990) 344}.

\bibitem{Bars:1991pt}
I.~Bars and K.~Sfetsos,
{\it Generalized Duality and Singular Strings in Higher Dimensions},
\doi{10.1142/S0217732392000963}{Mod. Phys. Lett. {\bf A7} (1992) 1091},
\arxiv{arXiv:hep-th/9110054}{arXiv:9110054 [hep-th]}.

\bibitem{Bars:1992ti}
I.~Bars and K.~Sfetsos,
{\it Global Analysis of New Gravitational Singularities in String and Particle Theories},
\doi{10.1103/PhysRevD.46.4495}{Phys. Rev.  {\bf D46} (1992) 4495},
\arxiv{arXiv:hep-th/9205037}{arXiv:9205037 [hep-th]}.

\bibitem{Gawedzki:1988nj}
K.~Gawedzki and A.~Kupiainen,
{\it Coset Construction from Functional Integrals},\hfill\break
\doi{10.1016/0550-3213(89)90015-1}{Nucl.\ Phys. {\bf B320} (1989) 625}.


\bibitem{Dunbar:1992gh}
D.~C.~Dunbar and K.~G.~Joshi,
{\it Characters for Coset Conformal Field Theories},\hfill\break
\doi{10.1142/S0217751X93001685}{Int.\ J.\ Mod.\ Phys.\ A {\bf 8} (1993) 4103},
\arxiv{arXiv:9210122}{arXiv:9210122 [hep-th]}.


\bibitem{Sfetsos:2010uq}
K.~Sfetsos and D.~C.~Thompson,
{\it On Non-Abelian T-Dual Geometries with Ramond Fluxes},
\doi{10.1016/j.nuclphysb.2010.12.013}{Nucl.\ Phys.\ B {\bf 846} (2011) 21},
\arxiv{arXiv:1012.1320}{arXiv:1012.1320 [hep-th]}.

\bibitem{Distler:1989xv}
  J.~Distler, Z.~Qiu,
  {\it BRS cohomology and a Feigin-Fuchs representation of Ka\v{c}-Moody and parafermionic theories},
  \doi{10.1016/0550-3213(90)90441-F}{Nucl.Phys. {\bf B336} (1990) 533-546}.


\end{thebibliography}
\end{document}